\documentclass[a4paper,11pt]{article}
\pdfoutput=1 
\usepackage{jinstpub} 

\usepackage{caption}
\usepackage[toc,page]{appendix}
\usepackage{hyperref}
\usepackage{lineno}
\usepackage{dcolumn}
\usepackage{bm}
\usepackage{multirow}
\usepackage{relsize}
\usepackage{environ}
\usepackage{placeins}
\usepackage{amsmath}

\usepackage{float}
\usepackage{xspace}
\usepackage[utf8]{inputenc}
\usepackage{listings}
\usepackage{xcolor}
\usepackage{enumitem}
\usepackage{todonotes}
\definecolor{codegreen}{rgb}{0,0.6,0}
\definecolor{codegray}{rgb}{0.5,0.5,0.5}
\definecolor{codepurple}{rgb}{0.58,0,0.82}
\definecolor{backcolour}{rgb}{0.95,0.95,0.92}
\lstdefinestyle{mystyle}{
    backgroundcolor=\color{backcolour},   
    commentstyle=\color{codegreen},
    keywordstyle=\color{magenta},
    numberstyle=\tiny\color{codegray},
    stringstyle=\color{codepurple},
    basicstyle=\ttfamily\scriptsize,
    breakatwhitespace=false,         
    breaklines=true,                 
    captionpos=b,                    
    keepspaces=true,                 
    numbers=left,                    
    numbersep=5pt,                  
    showspaces=false,                
    showstringspaces=false,
    showtabs=false,                  
    tabsize=1,
}

\newcommand{\Eqnref}[1]{Eq.~\ref{#1}}

\newcommand{\figref}[1]{Fig.~\ref{#1}}
\newcommand{\Figref}[1]{Figure~\ref{#1}}
\newcommand{\tabref}[1]{Tab.~\ref{#1}}
\newcommand{\Tabref}[1]{Table~\ref{#1}}

\newcommand{\Secref}[1]{Section~\ref{#1}}

\newcommand{\imagefolder}{./figures}
\sloppypar
                             
\title{Correction of the baseline fluctuations in the GEM-based ALICE TPC}
\abstract{
To operate the ALICE Time Projection Chamber in continuous mode during the Run~3 and Run~4 data-taking periods of the Large Hadron Collider, the multi-wire proportional chamber-based readout was replaced with gas-electron multipliers. As expected, the detector performance is affected by the so-called common-mode effect, which leads to significant baseline fluctuations. A detailed study of the pulse shape with the new readout has revealed that it is also affected by ion tails. Since reconstruction and data compression are performed fully online, these effects must be corrected at the hardware level in the FPGA-based common readout units. The characteristics of the common-mode effect and of the ion tail, as well as the algorithms developed for their online correction, are described in this paper. The common-mode dependencies are studied using machine-learning techniques. Toy Monte Carlo simulations are performed to illustrate the importance of online corrections and to investigate the performance of the developed algorithms.

}
\collaboration{ALICE TPC Collaboration}
\affiliation[1]{Bose Institute, Department of Physics  and Centre for Astroparticle Physics and Space Science (CAPSS), Kolkata, India}
\affiliation[2]{Comenius University Bratislava, Faculty of Mathematics, Physics and Informatics, Bratislava, Slovakia}
\affiliation[3]{Department of Physics, University of Oslo, Oslo, Norway}
\affiliation[4]{Department of Physics and Technology, University of Bergen, Bergen, Norway}
\affiliation[5]{European Organization for Nuclear Research (CERN), Geneva, Switzerland}
\affiliation[6]{Faculty of Engineering and Science, Western Norway University of Applied Sciences, Bergen, Norway}
\affiliation[7]{Helmholtz-Institut f\"{u}r Strahlen- und Kernphysik, Rheinische Friedrich-Wilhelms-Universit\"{a}t Bonn, Bonn, Germany}
\affiliation[8]{Helsinki Institute of Physics (HIP), Helsinki, Finland}
\affiliation[9]{High Energy Physics Group, Universidad Aut\'{o}noma de Puebla, Puebla, Mexico}
\affiliation[10]{Horia Hulubei National Institute of Physics and Nuclear Engineering, Bucharest, Romania}
\affiliation[11]{Indian Institute of Technology Bombay (IIT), Mumbai, India}
\affiliation[12]{Institut f\"{u}r Kernphysik, Johann Wolfgang Goethe-Universit\"{a}t Frankfurt, Frankfurt, Germany}
\affiliation[13]{Instituto de Ciencias Nucleares, Universidad Nacional Aut\'{o}noma de M\'{e}xico, Mexico City, Mexico}
\affiliation[14]{Joint Institute for Nuclear Research (JINR), Dubna, Russia}
\affiliation[15]{Lund University Department of Physics, Division of Particle Physics, Lund, Sweden}
\affiliation[16]{Nagasaki Institute of Applied Science, Nagasaki, Japan}
\affiliation[17]{Niels Bohr Institute, University of Copenhagen, Copenhagen, Denmark}
\affiliation[18]{NRNU Moscow Engineering Physics Institute, Moscow, Russia}
\affiliation[19]{Oak Ridge National Laboratory, Oak Ridge, Tennessee, USA}
\affiliation[20]{Physics department, Faculty of science, University of Zagreb, Zagreb, Croatia}
\affiliation[21]{Physikalisches Institut, Ruprecht-Karls-Universit\"{a}t Heidelberg, Heidelberg, Germany}
\affiliation[22]{Physik Department, Technische Universit\"{a}t M\"{u}nchen, Munich, Germany}
\affiliation[23]{PINSTECH, Islamabad, Pakistan}
\affiliation[24]{Research Division and ExtreMe Matter Institute EMMI, GSI Helmholtzzentrum f\"ur Schwerionenforschung GmbH, Darmstadt, Germany}
\affiliation[25]{The Henryk Niewodniczanski Institute of Nuclear Physics, Polish Academy of Sciences, Cracow, Poland}
\affiliation[26]{The University of Texas at Austin, Austin, Texas, USA}
\affiliation[27]{Universidade de S\~{a}o Paulo (USP), S\~{a}o Paulo, Brazil}
\affiliation[28]{University of Houston, Houston, Texas, USA}
\affiliation[29]{University of Jyv\"{a}skyl\"{a}, Jyv\"{a}skyl\"{a}, Finland}
\affiliation[30]{University of Tennessee, Knoxville, Tennessee, USA}
\affiliation[31]{University of Tokyo, Tokyo, Japan}
\affiliation[32]{Variable Energy Cyclotron Centre, Kolkata, India}
\affiliation[33]{Wayne State University, Detroit, Michigan, USA}
\affiliation[34]{Wigner Research Centre for Physics, Budapest, Hungary}
\affiliation[35]{Yale University, New Haven, Connecticut, USA}

\author[4]{J.~Alme,}
\author[12]{T.~Alt,}
\author[10]{C.~Andrei,}
\author[21]{V.~Anguelov,}
\author[12]{H.~Appelsh\"{a}user,}
\author[5,35,*]{M.~Arslandok, \note[*]{Corresponding authors}}\emailAdd{mesut.arslandok@cern.ch}

\author[34]{G.G.~Barnaf\"{o}ldi,}
\author[12]{E.~Bartsch,}
\author[21,24]{P.~Becht,}
\author[28]{R.~Bellwied,}
\author[18,21]{A.~Berdnikova,}
\author[12]{N.~Bialas,} 
\author[12]{P.~Bialas,} 
\author[1]{S.~Biswas,}
\author[21,24]{B.~Blidaru,}
\author[34]{L.~Boldizs\'{a}r,}
\author[12]{L.~Bratrud,}
\author[24]{P.~Braun-Munzinger,}
\author[27]{M.~Bregant,}
\author[19]{C.L.~Britton,} 
\author[21]{S.~Brucker,}
\author[8]{E.J.~Br\"{u}cken,} 
\author[12]{H.~B\"{u}sching,}

\author[9]{R.~Soto Camacho,} 
\author[30]{A.J.~Castro,} 
\author[21]{P.~Chatzidaki,}
\author[15]{P.~Christiansen,}
\author[19]{L.G.~Clonts,} 
\author[19,\dagger]{T.M.~Cormier, \note[$\dagger$]{Deceased}}

\author[11]{P.~Dhankher,}
\author[12]{S.~Dittrich,} 

\author[35]{R.~Ehlers,}
\author[19]{M.N.~Ericson,} 
\author[19]{N.B.~Ezell,}

\author[22]{L.~Fabbietti,}
\author[28]{F.~Flor,}

\author[17]{J.J.~Gaardh{\o}je,}
\author[27]{M.G.~Munhoz,}
\author[24]{C.~Garabatos,}
\author[22,24]{P.~Gasik,}
\author[12]{T.~Geiger,} 
\author[34]{\'{A}.~Gera,} 
\author[21]{P.~Gl\"{a}ssel,}
\author[16]{D.J.Q.~Goh,}
\author[12]{A.~Grein,}
\author[12]{T.~Gundem,}
\author[31]{T.~Gunji,}

\author[24]{M.~Habib,}
\author[16]{H.~Hamagaki,}
\author[34]{G.~Hamar,} 
\author[17]{J.C.~Hansen,}
\author[35]{J.W.~Harris,}
\author[7]{P.~Hauer,}
\author[31]{S.~Hayashi,}
\author[24]{J.~Hehner,} 
\author[8]{J.K.~Heino,}
\author[12]{E.~Hellb\"{a}r,}
\author[6]{H.~Helstrup,}
\author[12]{M.~Hemmer,}
\author[10]{A.~Herghelegiu,} 
\author[27]{R.A.~Hernandez,} 
\author[27]{H.D.~Hernandez Herrera,} 
\author[8]{T.E.~Hilden,}
\author[30]{C.~Hughes,}
\author[21]{S.~Hummel,}

\author[24,*]{M.~Ivanov,} \emailAdd{marian.ivanov@cern.ch}

\author[12]{J.~J\"{a}ger,}
\author[12]{J.~Jung,}
\author[12]{M.~Jung,}

\author[8]{E.~Kangasaho,}
\author[7]{B.~Ketzer,}
\author[12]{S.~Kirsch,}
\author[12]{M.~Kleiner,}
\author[22]{T.~Klemenz,}
\author[28]{A.G.~Knospe,}
\author[25]{M.~Kowalski,}
\author[24]{L.~Kreis,}
\author[12]{M.~Kr\"{u}ger,}
\author[21]{N.~Kupfer,}

\author[22]{R.~Lang,}
\author[22]{L.~Lautner,}
\author[22]{M.~Lesch,}
\author[5]{Y.~Lesenechal,} 
\author[12]{F.~Liebske,}
\author[24]{C.~Lippmann,}

\author[35,\dagger]{R.D.~Majka,}
\author[26]{C.~Markert,}
\author[27]{T.A.~Martins,} 
\author[24]{S.~Masciocchi,}
\author[15]{O.~Matonoha,}
\author[25]{A.~Matyja,}
\author[2]{M.~Meres,}
\author[22]{D.L.~Mihaylov,}
\author[24]{D.~Mi\'{s}kowiec,}
\author[21]{T.~Mittelstaedt,}
\author[22]{C.~Mordasini,}
\author[24]{T.~Morhardt,} 
\author[21]{S.~Muley,} 
\author[35]{J.~Mulligan,}
\author[12]{R.H.~Munzer,}
\author[31]{H.~Murakami,}
\author[7]{K.~M\"{u}nning,}

\author[15]{A.~Nassirpour,}
\author[30]{C.~Nattrass,}
\author[17]{B.S.~Nielsen,}
\author[27]{W.A.V.~Noije,}

\author[16]{M.~Ogino,}
\author[30]{A.C.~Oliveira Da Silva,}
\author[15]{A.~Oskarsson,} 
\author[16]{K.~Oyama,}
\author[15]{A.~\"Onnerstad,}
\author[15]{L.~\"Osterman,} 

\author[21]{Y.~Pachmayer,}
\author[13]{G.~Pai\'{c},}
\author[32]{R.N.~Patra,}
\author[12]{V.~Peskov,} 
\author[10]{M.~Petris,} 
\author[10]{M.~Petrovici,}
\author[20]{M.~Planinic,}
\author[10]{L.~Prodan,}

\author[10]{A.~Radu,} 
\author[19]{J.~Rasson,} 
\author[19]{K.F.~Read,}
\author[4]{A.~Rehman,}
\author[12]{R.~Renfordt,} 
\author[3]{K.~R{\o}ed,}
\author[4]{D.~R\"ohrich,}
\author[21]{E.~Rubio,} 
\author[19]{A.~Rusu,}

\author[27]{B.C.S.~Sanches,} 
\author[26]{J.~Schambach,}
\author[12]{S.~Scheid,}
\author[24]{C.~Schmidt,}
\author[30]{A.~Schmier,}
\author[24]{K.~Schweda,}
\author[31]{D.~Sekihata,}
\author[21]{S.~Siebig,} 
\author[27]{R.W.D.~Silva,} 
\author[15]{D.~Silvermyr,}
\author[2]{B.~Sitar,}
\author[35]{N.~Smirnov,}
\author[21]{H.K.~Soltveit,} 
\author[30]{S.P.~Sorensen,} 
\author[21]{J.~Stachel,}

\author[22]{L.~\v{S}erk\v{s}nyt\.{e},}

\author[4]{G.~Tambave,}

\author[4]{K.~Ullaland,}
\author[22]{B.~Ulukutlu,}

\author[34]{D.~Varga,}
\author[15]{O.~Vazquez Rueda,}
\author[4]{A.~Velure,} 
\author[9]{S.~Vergara Lim\'on,}
\author[21]{O.~Vorbach,} 
\author[24]{B.~Voss,}

\author[12]{C.~Weidlich,} 
\author[12]{J.~Wiechula,}
\author[21]{B.~Windelband,}
\author[22]{S.~Winkler} 

\arxivnumber{xxxxx} 

\begin{document}
\lstset{style=mystyle}
\maketitle

\section{Introduction} \label{sec:intro}

Charged particles passing through the active volume of the Time Projection Chamber (TPC) ionize the gas along their path. The electrons generated in this process drift toward the end plates on which the readout chambers are mounted. By amplifying the signal in the readout chambers, the TPC provides a three-dimensional reconstruction of the charged-particle tracks. In the data-taking periods Run~1 (2009--2013) and Run~2 (2015--2018) of the Large Hadron Collider (LHC), the readout chambers of the ALICE TPC~\cite{Alme:2010ke, ALICE:2000jwd} consisted of multiwire proportional chambers (MWPCs)~\cite{Rossegger:2010zz}. The ions generated during the amplification process in the MWPCs were blocked by a ``gating grid", a series of wires located between the cathode wires and the drift volume. However, the time requirements of the operation with gating grid limited the maximum readout rate of the TPC to around 3~kHz. On the other hand, operation of the MWPC-based TPC without ion gating would lead to intolerably large space-charge distortions in the drift region. Therefore, in order to operate the TPC with the expected minimum bias Pb--Pb collision rate of 50~kHz in Run~3 and Run~4 (2022--2030)~\cite{ALICE:2012dtf}, all readout chambers were replaced by gas electron multipliers (GEMs) during the Long Shutdown 2 (LS2) (December 2018 -- March 2022) of the CERN LHC, since stacks of GEMs have intrinsic ion blocking capabilities~\cite{Sauli:2016eeu}. To achieve the required gain while effectively suppressing the back-flow of ions, a stack of four GEM foils named from top to bottom (i.e. from drift to pad plane ): GEM1, GEM2, GEM3, and GEM4, was employed for all chambers~\cite{TDR:tpcUpgrade}. At the same time, due to the change from triggered to continuous operation, the front-end electronics (FEE) had to be replaced. A detailed description of the upgrade, from the design of the chambers and the new FEE to the pre-commissioning of the upgraded TPC, can be found in Ref.~\cite{ALICETPC:2020ann}.

\begin{figure}[h]
	\centering
	\includegraphics[width=0.4\linewidth]{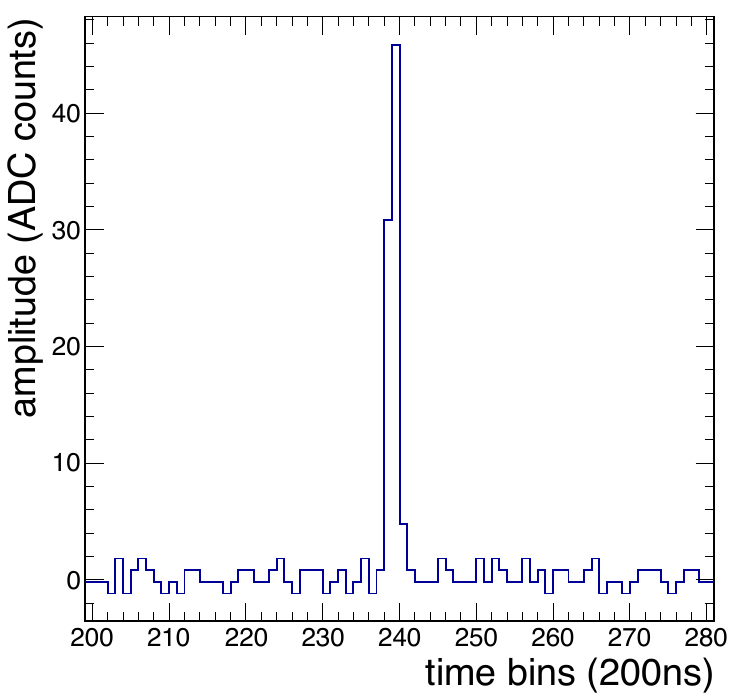}
	\caption{Typical electronics response of a single readout channel~\cite{ALICETPC:2020ann,TDR:tpcUpgrade}.}
	\label{fig:fig0_GEM_pulse}
 \end{figure}

The typical semi-Gaussian pulse shape of a single readout channel is shown in~\figref{fig:fig0_GEM_pulse}.
The signals from the charged particles are influenced by various effects, such as diffusion and track inclination angle. They also exhibit a characteristic undershoot, due to capacitive coupling across the induction gap between the pad plane and the GEM foils, and a long overshoot after the signal pulse caused by the slow movement of the ions in the induction gap. These two effects are referred to as ''common-mode" (CM) effect and ''ion tail" (IT), respectively. They are more prominent for high multiplicity environments and, if not accounted for, lead to significant deterioration of particle identification (PID) and tracking performance of the detector. Detailed studies conducted during Run~1 and Run~2, where similar baseline fluctuations were observed, emphasize the crucial need to understand, simulate, and correct for these effects to maintain the performance of the TPC~\cite{Arslandok:2022dyb}.

This paper is organized as follows; \Secref{Measurements} describes the measurements obtained during the pre-commissioning phase during LS2. These data served as input for the initial analysis of the two effects. In \Secref{cmeffect}, the dependencies of the common-mode effect are investigated using machine learning techniques. In \Secref{iontail}, a detailed analysis of the ion-tail properties is performed. In \Secref{corrections}, the two online correction algorithms as implemented at the hardware level in the FPGA-based common readout units (CRUs) are described, and a proof of principle is provided. Toy Monte Carlo (MC) simulations are performed to demonstrate the impact of the two effects on the TPC signals and to investigate the performance of the two online correction algorithms. 

\section{Measurements} \label{Measurements}

\begin{figure}[h]
	\centering
	\includegraphics[width=0.45\linewidth]{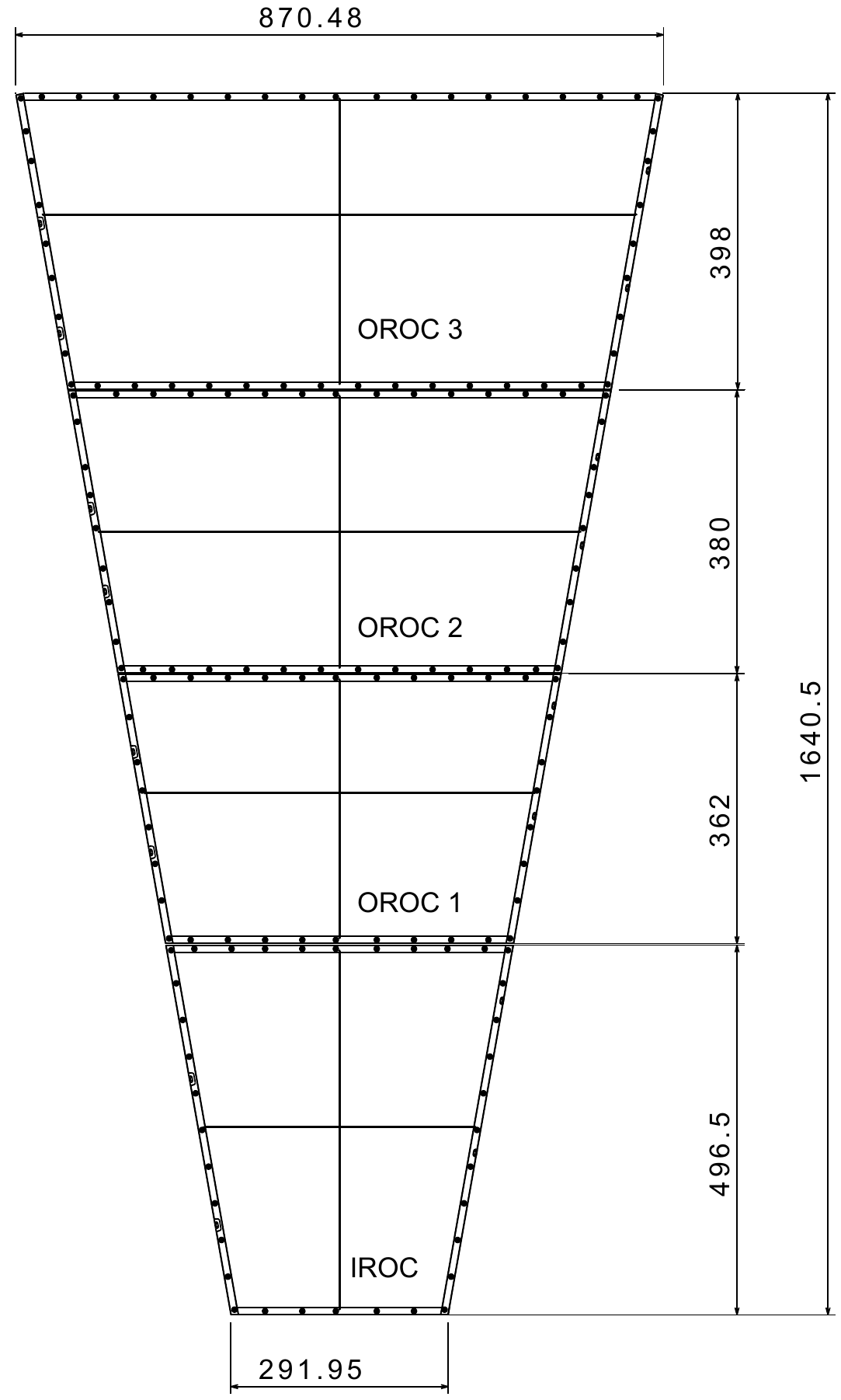}
	\caption{Dimensions (mm) of a sector of the TPC with four GEM stacks. The spacer cross, a structure that ensures the mechanical stability of the GEM foils against electrostatic forces, is shown as two 1.5 mm wide and 2 mm deep bars in the longitudinal and transverse directions for each stack~\cite{ALICETPC:2020ann}.}
	\label{fig:GEM_segmentation}
 \end{figure}

A quadruple GEM configuration with foils having different hole pitch was optimized for the ALICE TPC during an extensive R\&D phase. The TPC volume is divided equally into two readout sides (A-side and C-side) by a central electrode. The TPC sectors are positioned on the endplates, each covering 20$^{\circ}$ in azimuth. They are radially segmented into an inner and outer readout chambers (IROC and OROC, respectively). The OROC is further subdivided into three individual GEM stacks, therefore a TPC sector consists of a total of four GEM stacks, labeled IROC (stack~0), OROC~1 (stack~1), OROC~2 (stack~2), and OROC~3 (stack~3) (see \figref{fig:GEM_segmentation}). The pad signals are read out by the FEE mounted on the TPC end plates at a sampling rate of 5~MHz, which corresponds to a time bin duration of 200~ns~\cite{Appelshaeuser:2231785,ALICETPC:2020ann}. Custom-made ASICs, called ``SAMPA" chips~\cite{Hernandez:2019pqz}, are responsible for the signal amplification, shaping, and analog-to-digital conversion. The digitized data are then streamed to the CRUs, where the baseline subtraction, common-mode effect correction, ion-tail filtering, and zero suppression (ZS) are performed. 

\subsection{Laser data}
 To study the common-mode and ion-tail effects, the laser calibration system of the TPC was used~\cite{Renault:2005tr, Borge:laser}. Laser data were collected for all sectors during the pre-commissioning of the TPC in the clean room located at the LHC Point~2~\cite{ALICETPC:2020ann}. The laser is a pulsed (5~ns pulse duration, 10~Hz repetition rate) Nd:YAG laser equipped with two frequency doublers, resulting in a final wavelength of 266~nm, which corresponds to an ionization energy of 4.66~eV. 
 However, the ionization potential of the TPC gas components is much larger ($E_{\mathrm Ne}\approx22$~eV, $E_{\mathrm CO_{2}}\approx14$~eV, $E_{\mathrm N_{2}}\approx16$~eV). Therefore, with two-photon processes~\cite{Hilke:1986bw}, the laser ionizes organic impurities (approximately 1~ppm) in the gas with ionization potentials of $5-8$~eV.
  \begin{figure}[h]
	\centering
	\includegraphics[width=0.6\linewidth]{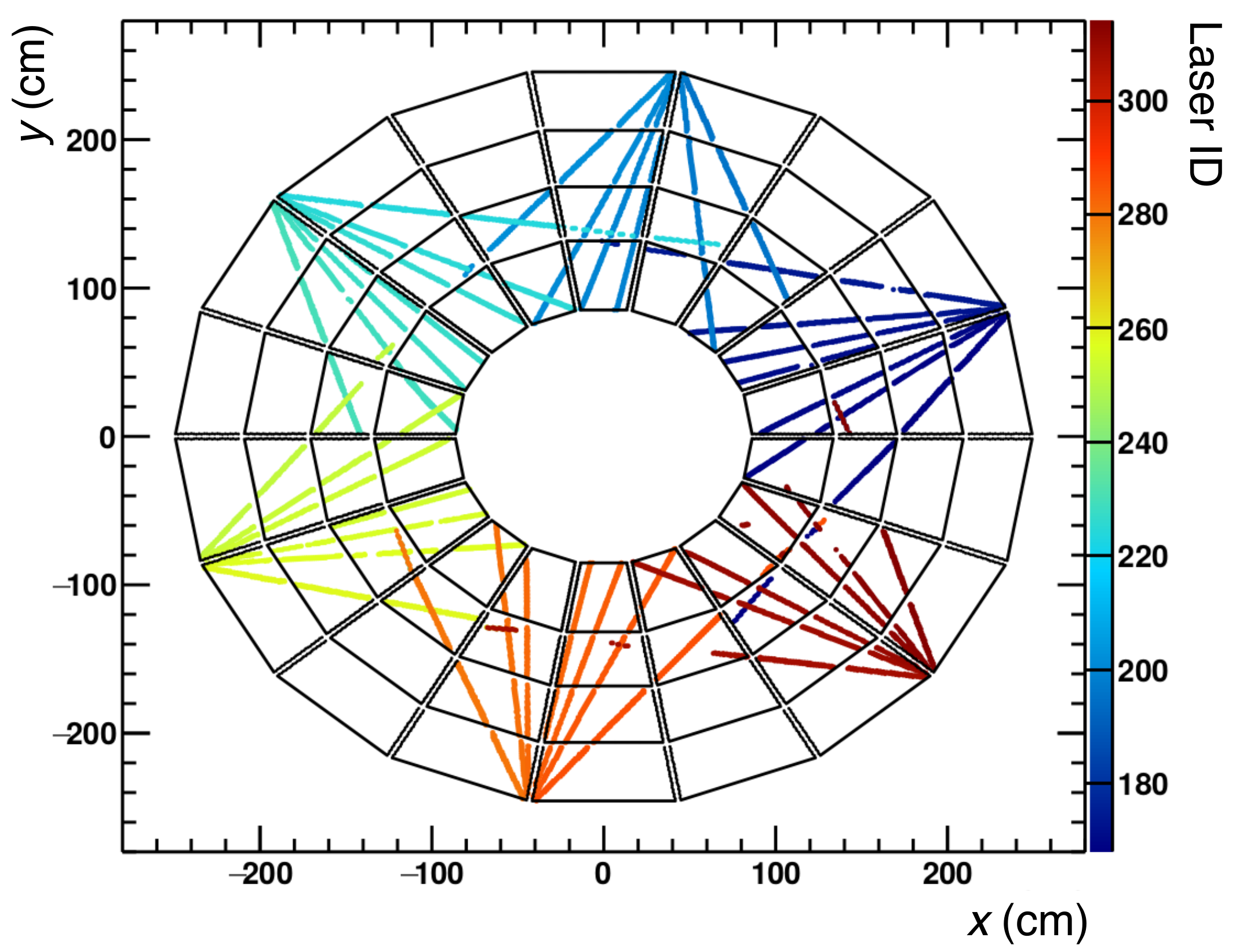}
	\caption{Reconstructed laser tracks of a particular bundle on the C-side of the TPC. The black lines indicate the sector boundaries. The \textit{laser ID} is a unique number assigned to each of the 336 laser tracks.}
	\label{fig:laserSys}
 \end{figure}
Through a series of mirrors, beam splitters, prisms, and micromirror bundles, 336 laser tracks simultaneously irradiate the TPC. In each TPC half, 6 wide laser beams illuminate 24 bundles of micromirrors located at four nearly equidistant positions along the LHC beam direction. For each bundle, 7 narrow reflected beams enter the TPC gas parallel to the end plates. \Figref{fig:laserSys} shows the reconstructed laser tracks of the C-side bundle 0 (the bundle closest to the endplate), with the color scale indicating the \textit{laser ID}.

\begin{figure}[h]
    \centering 
    \includegraphics[width=0.7\textwidth]{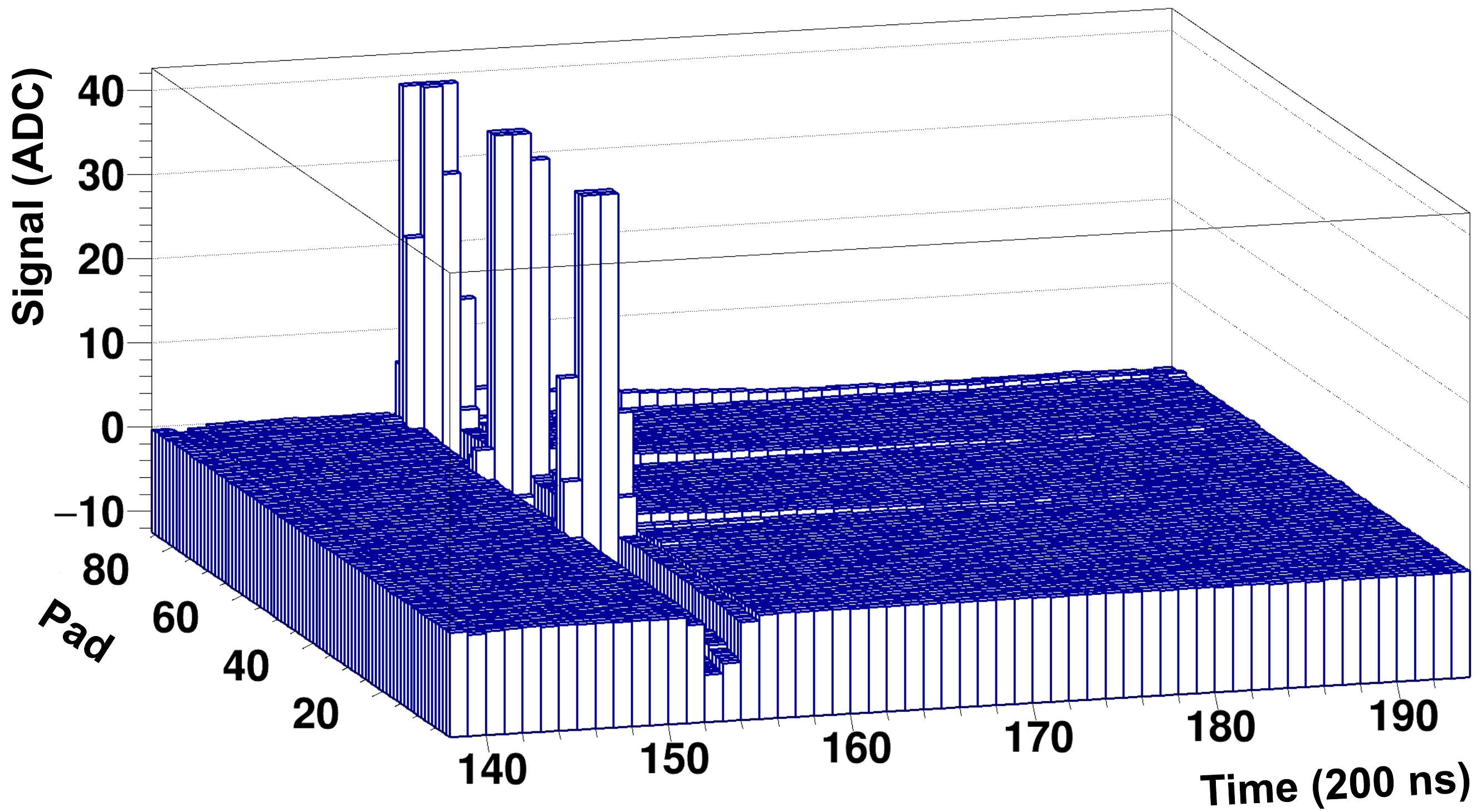}
    \caption{Laser signals and induced common-mode signals in the pads of a given pad row. The signal height axis is zoomed-in such that the signals are not to scale. The ion tail is also visible for the signal pads.} 
    \label{fig:cm_effect}
\end{figure}

In these measurements, for each laser event (corresponding to one laser pulse), raw data with a length of about 500 time bins (one time bin corresponds to 200~ns) were streamed from the FEE to the CRUs, corresponding to the full electron drift time. The data taking was triggered at 10~Hz by the laser system. For each TPC sector, a sufficient amount of laser events (approximately 400--1200) were collected. To increase the signal-to-noise ratio, the signals were averaged over all available events in order to reach better precision of the shapes of the common-mode signal and of the ion tail. \Figref{fig:cm_effect} shows three laser clusters\footnote{A cluster is defined as concentrated deposited charge detected within a search window of 3 bins in pad direction and 3 bins in time direction.} detected on one TPC pad row. The simultaneous common-mode signal is seen as an undershoot in the remaining pads, which are referred to as \textit{non-signal} or \textit{empty} pads. After the signal pulse, the long ion tail is also clearly visible in the signal pads. 

\begin{figure}[h]
    \centering
    \includegraphics[width=0.95\textwidth]{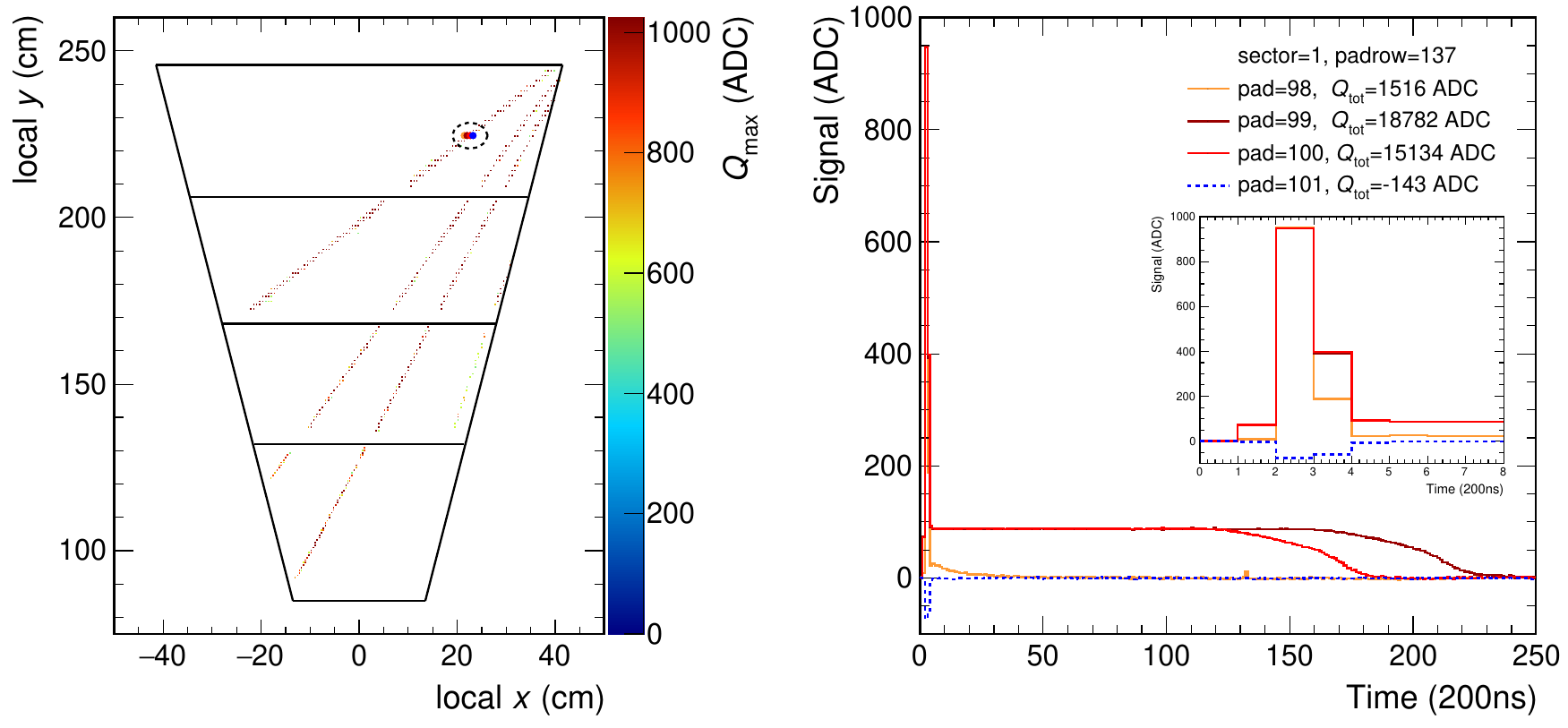}
    \caption{Left: four laser tracks projected on to the local x--y plane. The dashed oval highlights saturated signals. Right: saturated laser signals of the cluster highlighted in the left panel.} 
    \label{fig:laser_saturated}
\end{figure}

Very large signals saturating the dynamic range of the SAMPA chip were discarded from the analysis, since their actual amplitude is not known. Moreover, very large charges fed into the input of the SAMPA chip may lead to loss of sensitivity of this channel for a short time.
This can be seen as a saturation of the SAMPA response at a constant value of about 100~ADC, as shown in the right panel of \figref{fig:laser_saturated} with red and brown solid lines. Note that this effect is an artifact of the electronics and therefore does not affect the neighboring pads.

\subsection{Calibration pulser data} \label{pulser}
\begin{figure}[htbp]
	\centering
	\includegraphics[width=0.65\linewidth]{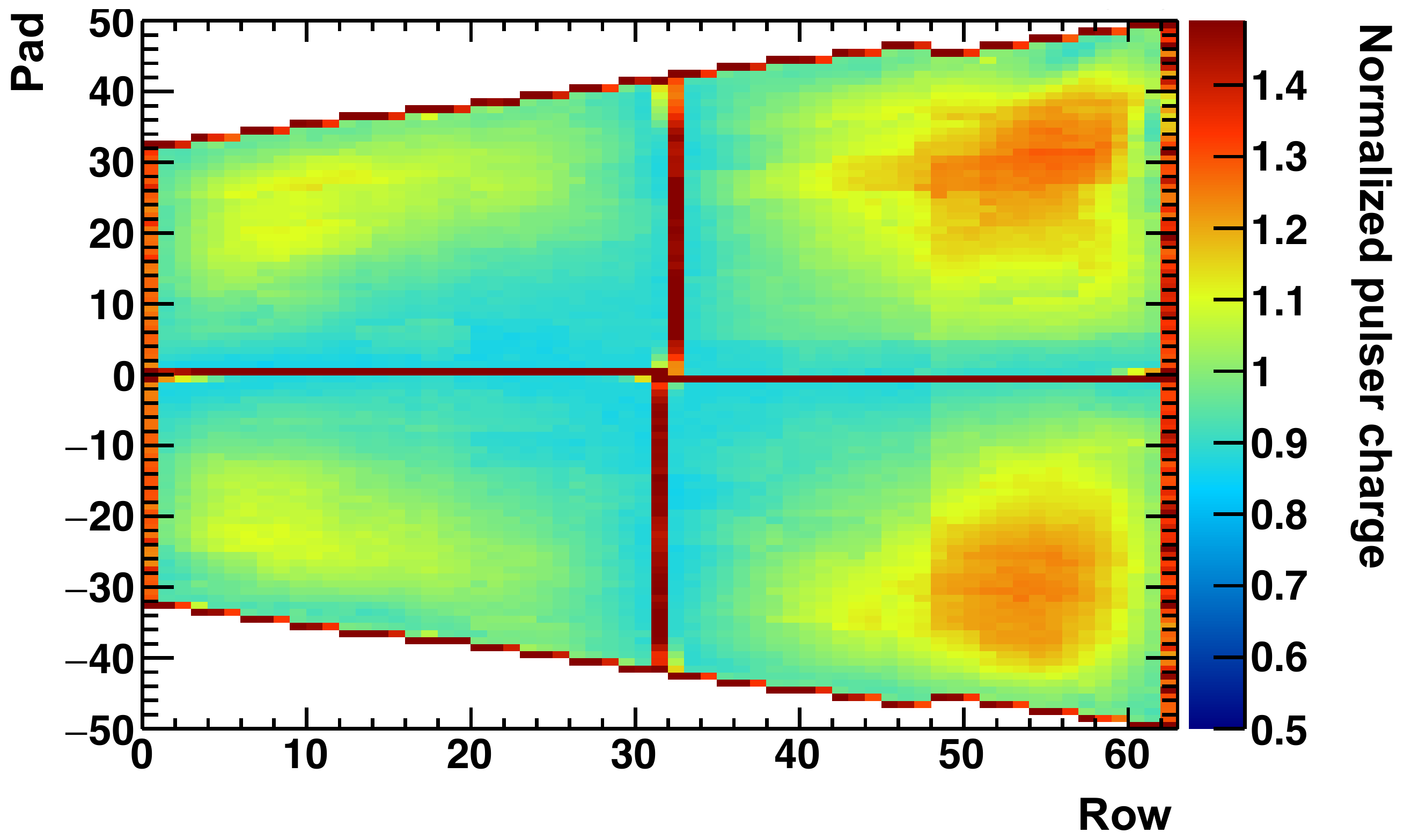}
	\caption{Ratio of pulser charge to median pulser charge in the stack for an IROC.}
	\label{fig:pulser}
\end{figure}

Measurements with a calibration pulser system were performed independently during the pre-commissioning phase in order to investigate the shaping characteristics of the FEE. These measurements involve injecting a pulse into the bottom electrode of the GEM4 foil (GEM4B), which induces a signal at the pad plane due to capacitive coupling. Ideally, the same charge should be measured in all pads of the same stack since they have the same dimensions. However, as shown in~\figref{fig:pulser}, significant variations are observed from pad to pad due to sagging of the GEM foil. The figure shows the normalized pulser charge, i.e., the charge normalized to the median charge in the stack, for an IROC. For pads positioned at the stack edges and under the spacer cross, the measured charge is significantly higher due to the presence of a dielectric material at these positions. Despite the spacer cross, the GEM4B bends slightly in the direction of the pad plane due to the stretching of the foil and electrostatic forces. This increases the capacitance and thus the pulser charge. As shown in~\figref{fig:pulser}, the relative change in the capacitance for a given pad reaches up to 50\%. 

The measured pulser charge is proportional to the pad capacitance, which is an essential parameter in the analysis of the common-mode effect. Since both, the pulser charge and the effective field between the GEM4B and the pad plane, are inversely proportional to the distance between the GEM4B and the pad plane (except for pads in the spacer cross and at the edges), the pulser charge is also a relevant parameter for the ion-tail studies.

\section{Common-mode effect analysis} \label{cmeffect}
\subsection{The common-mode effect}
The common-mode effect results from the capacitive coupling between the GEM foils and the pad plane. When electrons moving from GEM4 to the pad plane induce a signal on one pad, a simultaneous signal of opposite polarity (also called \textit{undershoot}) is induced on all pads of the same stack, as shown in~\figref{fig:cm_effect}. This is caused by a voltage drop across the GEM electrode due to the currents caused by the charges drifting through the induction gap. The magnitude of the undershoot in each pad for a given time bin, $Q_{\mathrm{pad}}^{\mathrm{ CM }}(t)$, is proportional to the sum of the positive signal in the stack at the same time bin. This pad- and time-independent proportionality factor will be referred to as \textit{common-mode fraction (CF) factor}, $k_{\mathrm{ CF, pad}}$, and is defined as
\begin{align}
k_{\mathrm{CF, pad}} = \frac{Q_{\mathrm{pad}}^{\mathrm{CM}}(t)}{\hspace{0.7cm} \left<Q_{\mathrm{pos}}(t)\right>_{\mathrm{stack}} \hspace{0.1cm}},
\label{eq:cf0}
\end{align}
where 
\begin{align}
\left<Q_{\mathrm{pos}}(t)\right>_{\mathrm{stack}} =\sum_{\substack{Q(t) > 0 \\ \mathrm{in stack}}} Q(t)\Big/N
\label{eq:avgsig0}
\end{align}
is the average positive signal in the stack and $N$ is the number of pads in the stack. The main objective of the common-mode effect analysis is to investigate the dependencies of the $k_{\mathrm{CF, pad}}$. The laser signals usually span three time bins (see \figref{fig:cm_effect}), so for the analysis of the effect, the common-mode charge and the average positive signal in the stack were summed over three time bins around the laser signal. Since the common-mode effect influences all pads in a given stack, the common-mode signals are also present in the laser signal pads, i.e., the true laser signal is slightly larger than the measured signal. This is taken into account in the analysis.

\subsection{Dependencies of the common-mode effect}
All possible dependencies of $k_{\mathrm{CF, pad}}$ were explored using the Random Forest (RF) machine learning algorithm~\cite{ho1995random} as implemented in the ROOTInteractive framework~\cite{rootInteractive}. To estimate the common-mode signal, 33\% of the non-signal pads were randomly selected as training data. Then, the selected sample was randomly subdivided into 200 estimators, and a decision tree with a depth of 12 was generated for each estimator. The dependencies of the $k_{\mathrm{CF, pad}}$ on all available variables were tested. The importance of each variable is listed in \Tabref{tab:importance}.

\begin{table}[h]
	\centering
	\begin{tabular}{cc}
		\hline
		Variable                                                            & Variable importance (\%)  \\
		\hline
		normalized pulser charge   ($Q^{\mathrm{norm}}_{\mathrm{pulser, pad}}$)  & 61.1                      \\
		stack type                                                          & 36.1                      \\
		average positive signal in stack                                    & 1.0                       \\
		fraction of signal pads in stack                                    & 0.8                       \\
	remaining dependencies & 1.0 \\	
		\hline    
	\end{tabular}
	\caption{Variable importance for $k_{\mathrm{CF, pad}}$, reflecting how many times the decision tree was divided because of that specific variable.}
	\label{tab:importance}
\end{table}
\begin{figure}[h]
  \centering 
  \includegraphics[width=\textwidth]{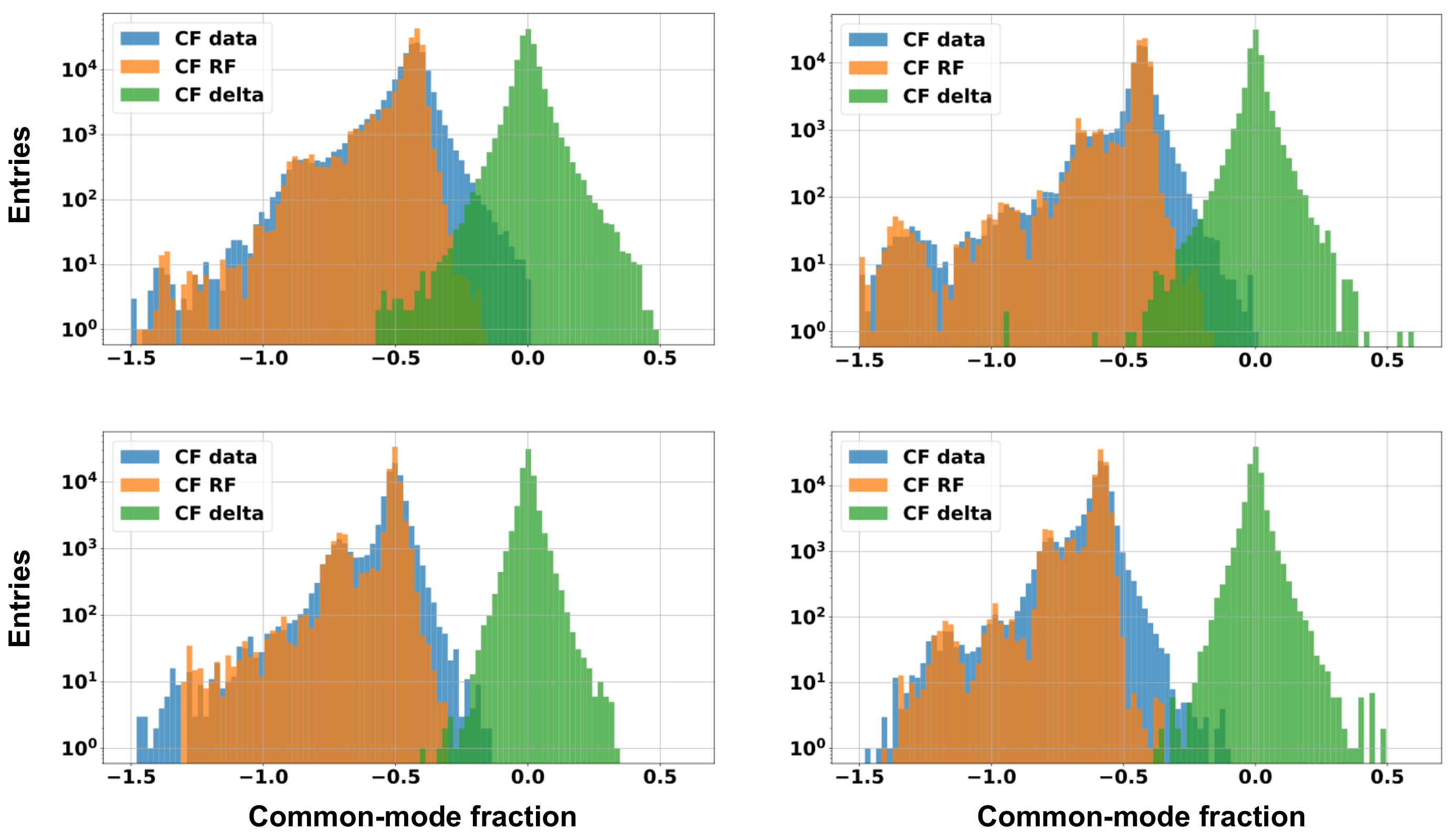}
  \caption{(Color online) The common-mode fraction obtained from data (blue), random-forest prediction (orange) and difference between the two (green), for IROC (top left), OROC1 (top right), OROC2 (bottom left) and OROC3 (bottom right). The data shown are those used for the random forest training.}
  \label{fig:trained_log}
\end{figure}
The normalized pulser charge, $Q^{\mathrm{norm}}_{\mathrm{pulser, pad}}$ (i.e., the pulser charge measured in a given pad normalized to the mean pulser charge in the stack), measured with the calibration pulser system and the stack type account for approximately 97\% of the dependencies. The $Q^{\mathrm{norm}}_{\mathrm{pulser, pad}}$ is responsible for the pad-by-pad capacitance variations, while the stack type (IROC, OROC1, OROC2, OROC3) accounts for the absolute stack capacitance due to the different stack dimensions. Note that the amplitude of the measured laser signal is reduced due to the underlying common-mode signal. This is referred to as missing charge, which is responsible for the second-order effects mentioned in \Tabref{tab:importance}; the \textit{average positive signal in the stack} and the \textit{fraction of signal pads in the stack}. The contribution of track-related properties such as the bundle (i.e. diffusion) and the beam (i.e. track inclination) was investigated, but no significant dependence was found. 
\begin{table}[h!]
\centering
\begin{tabular}{c|cc}
       & \begin{tabular}[c]{@{}c@{}} Stack area (mm$^2$) \end{tabular} & Peak position of CF  \\ \hline
IROC           & 171154                                                         & -0.42 ± 0.03\\
OROC1           & 174853                                                         & -0.43 ± 0.02\\
OROC2           & 231284                                                         & -0.50 ± 0.02 \\
OROC3           & 294836                                                         & -0.58 ± 0.02 \\ \hline
\end{tabular}
\caption{Position of the maximum of the CF distributions in \figref{fig:trained_log} for each stack type.}
\label{tab:CFmax}
\end{table} 

\begin{figure}[h]
    \centering 
    \includegraphics[width=0.7\textwidth]{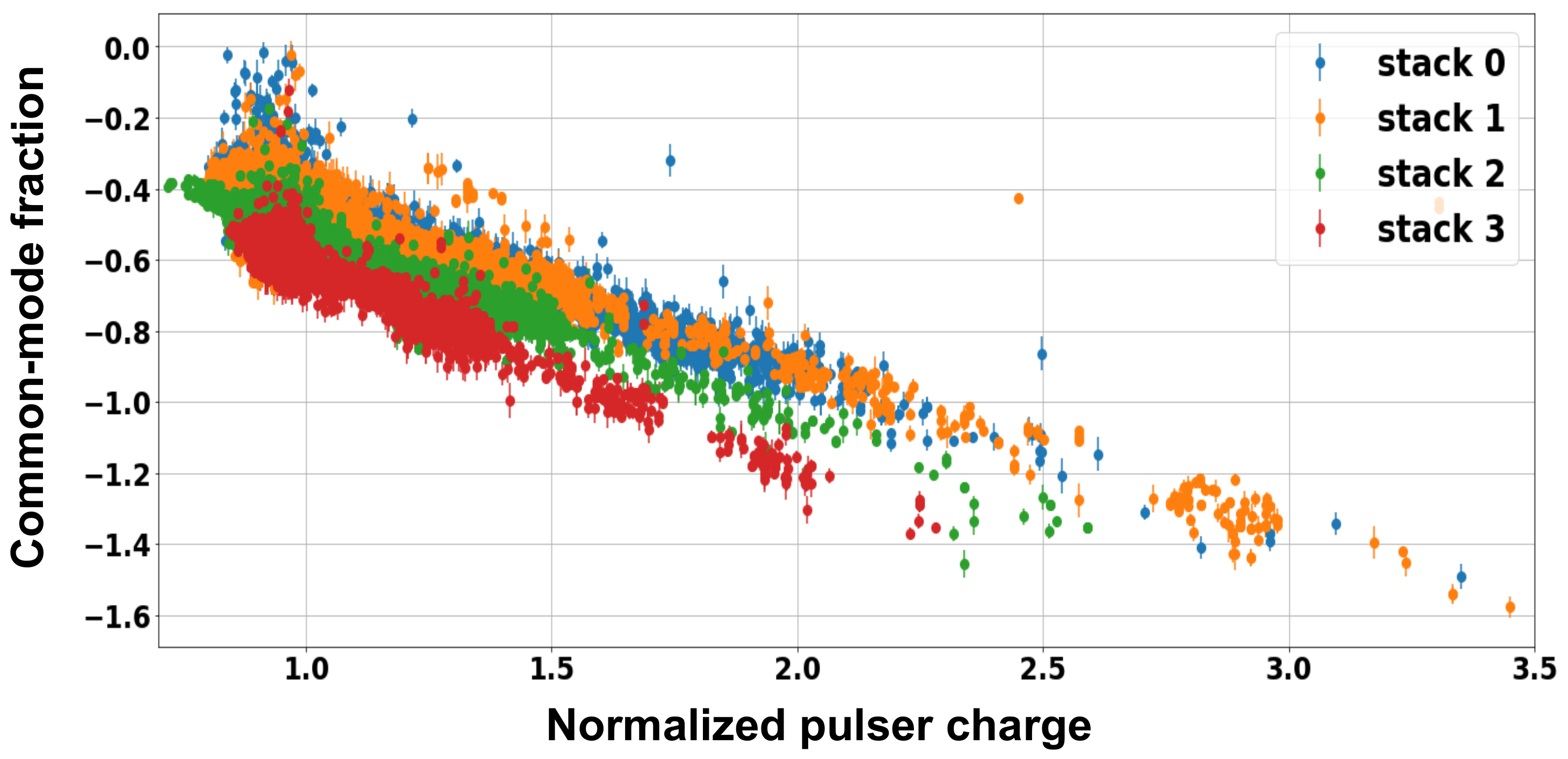}
    \caption{(Color online) The common-mode fraction ($k_{\mathrm{CF, pad}}$) as a function of normalized pulser charge ($Q^{\mathrm{norm}}_{\mathrm{pulser, pad}}$) for each stack, only for the data with uncertainty of $k_{\mathrm{CF, pad}}$ less than $5\%$.}
    \label{fig:cf_Qpulser}
\end{figure}

In \figref{fig:trained_log}, the $k_{\mathrm{CF, pad}}$ data, the RF prediction, and the difference between the two for each pad are plotted for the training data, where very good agreement between the data and the prediction is observed. The peak value of $k_{\mathrm{CF, pad}}$ moves towards larger (absolute) values as the stack area and hence capacitance increases (see \tabref{tab:CFmax}). The spread within each stack results from the pad-by-pad capacitance variations, which can be seen as a linear dependence between $k_{\mathrm{CF, pad}}$ and $Q^{\mathrm{norm}}_{\mathrm{pulser, pad}}$ in \figref{fig:cf_Qpulser}. The proportionality also holds for pads with much larger capacitance located at the chamber edges and crosses. By performing linear fits on the data points, the common-mode fraction of a pad can be expressed as
\begin{align}
k_{\mathrm{CF, pad}} = k_{\mathrm{stack}} \cdot Q^{\mathrm{norm}}_{\mathrm{pulser, pad}},
\label{eq:cf1}
\end{align}
where $k_{\mathrm{stack}}$ is the absolute value of the slope (0.44--0.58, depending on the stack type).

The following conclusion can be drawn from the studies: The $k_{\mathrm{CF, pad}}$ of a given pad depends mainly on the capacitance between the pad and the GEM stack (see \Tabref{tab:importance}), which can be described by the stack type and the normalized pulser charge, the latter reflecting the pad-by-pad capacitance variations within a given stack. 
Note that the missing charge has a negligible effect on the common-mode correction, as discussed above. However, it is accounted for in the full MC simulations that use a GEANT3 implementation of the TPC detector setup~\cite{TDR:O2}.

\section{Ion-tail analysis} \label{iontail}
\subsection{Ion tail in GEMs} \label{iontail1}
\begin{figure}[h]
    \centering 
    \includegraphics[width=\textwidth]{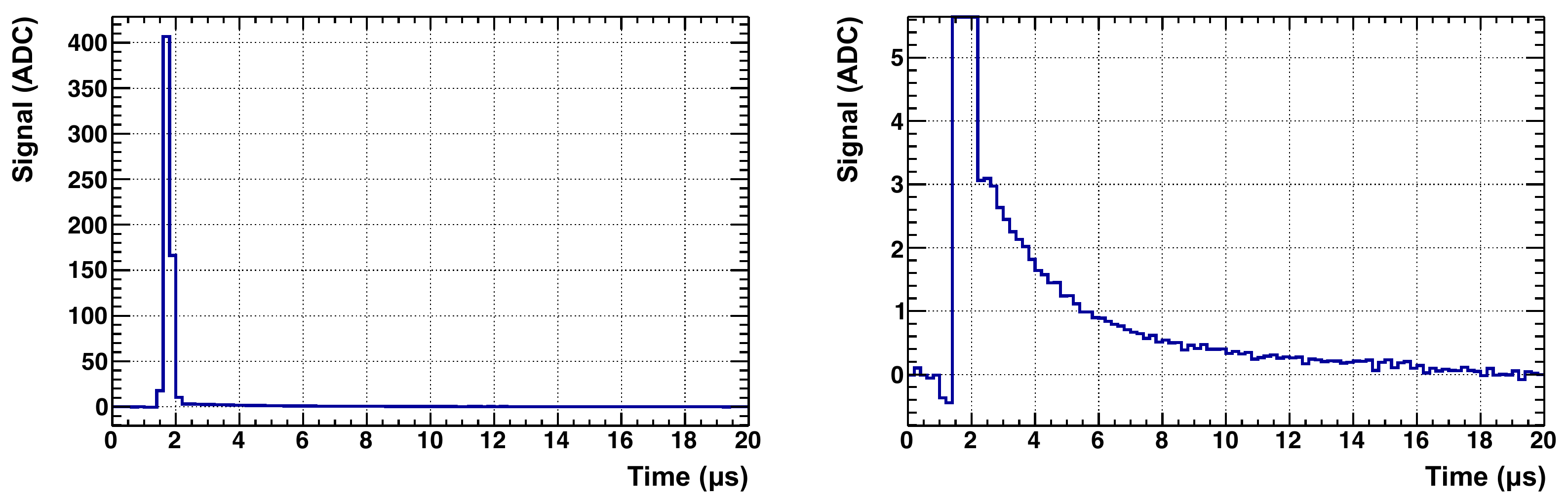}
    \caption{Left: Response of a pad to a laser signal. Right: The same signal zoomed in on the ($y$) axis. The undershoot observed before the signal pulse is the common-mode response due to signals in other pads of the same stack.}
    \label{fig:ionTail_sector_19_row_100_pad63_both}
\end{figure}

The analysis of laser data for the common-mode effect revealed unanticipated signal tails. In \figref{fig:ionTail_sector_19_row_100_pad63_both} (left) the response of a single pad to a laser pulse is shown. On the right, the same response is zoomed in on the signal ($y$) axis. After the signal peak, a long tail is observed, caused by the ions generated in the amplification process, lasting about 16~$\mu$s. In this example, the maximum of the tail is about 0.7\% of the maximum of the electron signal, while the integral of the tail, $Q_{\mathrm{tot}}^{\mathrm{tail}}$, corresponds to approximately 9\% of the total electron signal, $Q_{\mathrm{tot}}^{\mathrm{signal}}$. The shape and duration of the tail depend on the distance between the GEM4B and the pad plane, the distance of the signal from the center of gravity (COG) of the cluster along the pad direction, the track inclination, and the diffusion. Consequently, the above ratios are strongly influenced by these parameters.

A long ion tail as a negative undershoot was also observed during Run~1 and Run~2, where the TPC was based on the MWPC technology. The integral of the ion tail accounted for approximately 50\% of the total signal, resulting in significant degradation of the detector performance, especially in the case of out-of-bunch pile-up events~\cite{Arslandok:2022dyb}. Despite its positive nature and smaller amplitude, the ion tail still requires a correction during the Run 3 data-taking period, where, on average, five pile-up events are expected within a full drift time.

Simulations were performed to understand the origin of the ion tail observed in GEMs. For these simulations, only the last GEM foil (GEM4) was modeled, which is sufficient to a first approximation due to the shielding of the ions generated in the previous amplification stages by the GEM4 electrodes. The electric field maps were calculated using the finite element method as implemented in ANSYS~\cite{ANSYS}. The transport properties of the charge carriers were determined using Magboltz~\cite{Magboltz}, and their multiplication was simulated in Garfield++~\cite{garfieldpp}. The simulations show that two categories of ions contribute to the ion tail, which can be classified by their point of origin. They are created either in the GEM4 holes or in the induction gap, i.e., the region between GEM4B and the pad plane. While the first category can not be avoided and will be present in any GEM system, the latter is particular for the HV settings chosen for the ALICE TPC GEMs, where a high induction field (electric field between the GEM4B electrode and the pad plane, $E_{\mathrm{ind}}$) plays an important role in minimising the ion backflow. A detailed description of the HV settings can be found in~\cite{ALICETPC:2020ann}.
\begin{figure}[h]
    \centering 
    \includegraphics[width=0.65\textwidth]{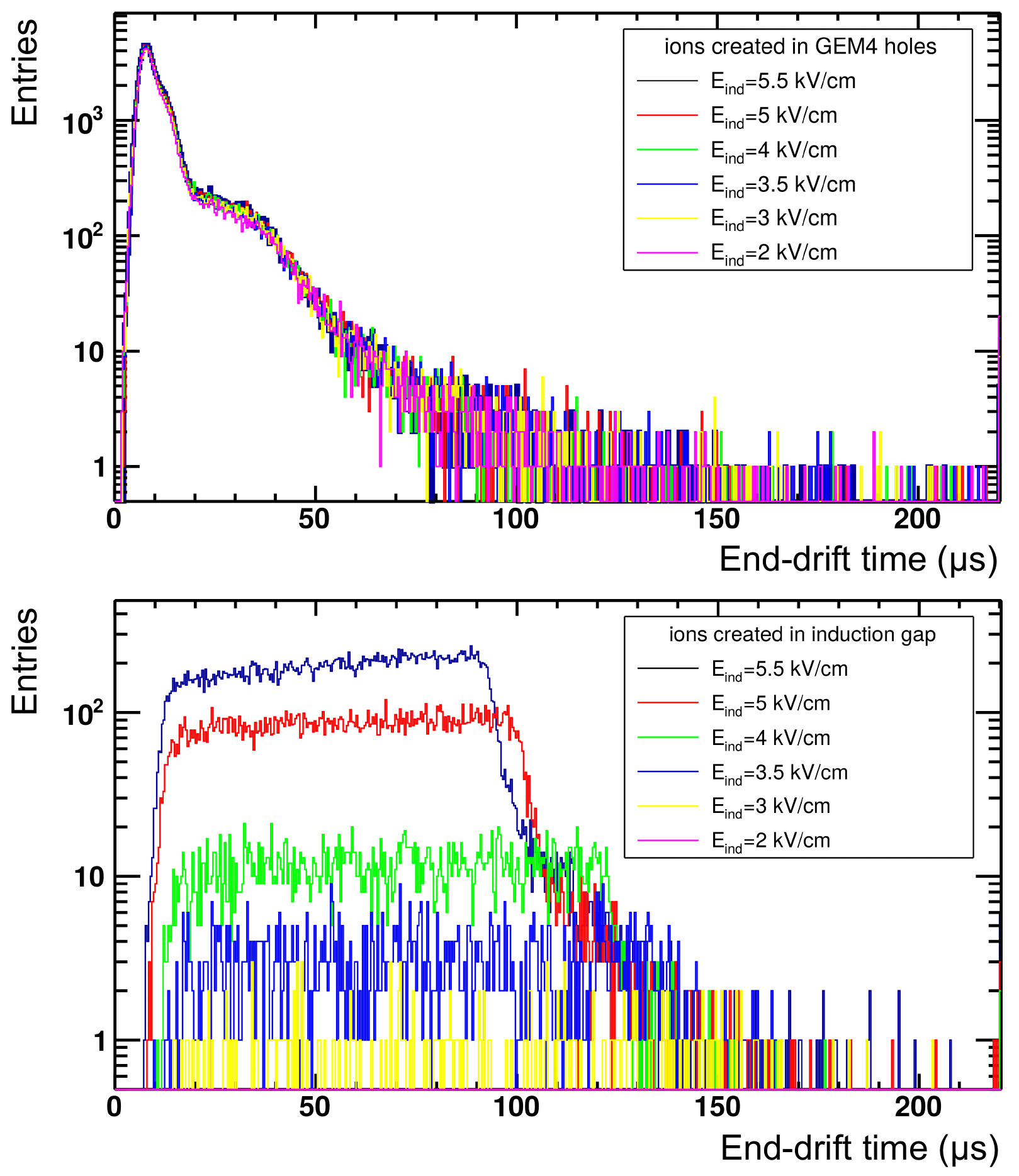}
    \caption{(Color online) Simulated drift times of ions until they reach an electrode for two categories of ions: those generated in the GEM4 holes (top) and in the induction gap (bottom). Different induction field values are shown as different colors, with the nominal value corresponding to 3.5~kV/cm.}
	\label{fig:garfTaku}
\end{figure} 

The simulation results are summarized in \figref{fig:garfTaku}, where the time between the creation of each ion until it is collected at an electrode or leaving the amplification area (the \textit{end-drift time}) is plotted separately for the two types of ions mentioned above. It can be seen that the number of ions generated in the amplification stage (top panel) is considerably larger than those generated in the induction gap (bottom panel). The former are referred to as the \textit{fast component} of the ion tail due to their sharp distribution in small end-drift time values, while the latter are referred to as \textit{slow component} due to their flat distribution. The distribution of the fast component does not depend on $E_{\mathrm{ind}}$, however, in the case of the slow component, the distribution becomes narrower with increasing $E_{\mathrm{ind}}$ and acquires a slight slope. Since the probability of ionization depends on the induction field, the number of produced ions rapidly decreases with decreasing $E_{\mathrm{ind}}$. As seen in the bottom panel of \figref{fig:garfTaku}, for the low $E_{\mathrm{ind}}$ settings, the end-drift time distribution of the slow component ions is flat, indicating a uniform production of electron/ion pairs in the induction gap. On the other hand, for higher $E_{\mathrm{ind}}$ and in particular for $E_{\mathrm{ind}} > 4$~kV/cm (see Fig.~2 of Ref~\cite{Ball:2014qaa}), the electron/ion pair creation probability is larger closer to the pad plane, due to avalanche effects. Note that the nominal induction field value, $E^{100\%}_{\mathrm{ind}}$, is set to 3.5~kV/cm.

\subsection{Induction field dependence} 

A dedicated set of measurements was added to the pre-commissioning program to study the tail properties and disentangle the contributions of the two different ion types. In these measurements, 5000 laser events were collected for two TPC sectors. The value of the induction field was set to 50\% of the nominal value and then increased in 5\% steps, from 75\% to 100\%. The high number of laser events compared to the standard laser calibration runs (with 1000 events) was chosen to ensure a good signal-to-noise ratio, since the tail magnitude is relatively small.

\begin{figure}[h]
	\centering
	\includegraphics[width=0.6\linewidth]{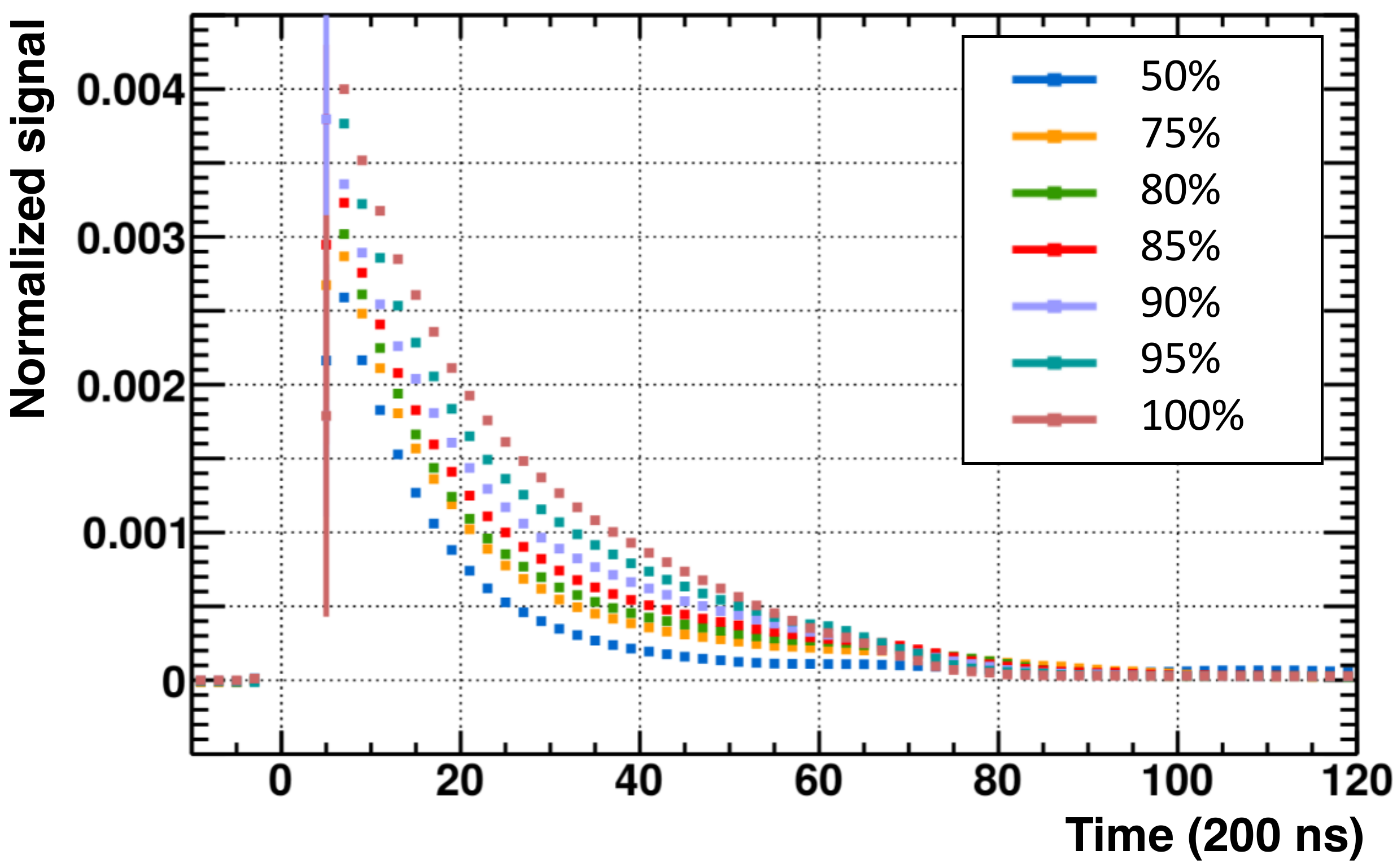}
    \caption{(Color online) Normalized signals as a function of time for the different induction field settings, averaged over all central pads (within 0.2 cm from cluster COG). Numbers listed in the legend correspond to percentages of $E^{100\%}_{\mathrm{ind}}$. The error bars (standard error of the mean) are smaller than the marker size.}
	\label{fig:EindScanPlot}
\end{figure}

\Figref{fig:EindScanPlot} shows the average normalized laser signals of all central pads. The normalization is with respect to the maximum value of a given pad signal. It demonstrates that the magnitude of the tail decreases with decreasing induction value and the shape also varies depending on the induction field value. Apart from the magnitude, the shape of the tail is also different. In particular, the tail decays faster for lower values of the induction field. Note that, due to the steeply rising nature of the tail, as illustrated in~\figref{fig:ionTail_sector_19_row_100_pad63_both}, and the fluctuations in the laser signal positions, the bins near the peak exhibit greater fluctuations, leading to increased statistical errors.

\subsection{Estimation of the contribution from the two categories of ions} \label{itcontr}

Based on the simulation results, an estimate of the contributions from the two categories of ions described in \Secref{iontail1} can be made. It can be assumed that for $E^{50\%}_{\mathrm{ind}} = 1.75$~kV/cm the contribution of the slow component is negligible with respect to higher field settings as see in~\figref{fig:garfTaku}. Since the fast component is practically independent of the induction field, the slow component contribution for a given $E_{\mathrm{ind}}$ value can be estimated as the difference between the value with this induction field and the one with $E^{50\%}_{\mathrm{ind}}$, i.e., one can write
\begin{align}
Q_{\mathrm{fast}}(E_{\mathrm{ind}}) & \approx Q_{\mathrm{meas}}(E^{50\%}_{\mathrm{ind}}) \\[5pt]
Q_{\mathrm{slow}}(E_{\mathrm{ind}}) & \approx Q_{\mathrm{meas}}(E_{\mathrm{ind}}) -  Q_{\mathrm{meas}}(E^{50\%}_{\mathrm{ind}}).
\label{eq:components}
\end{align}

\begin{figure}[h]
	\centering
	\includegraphics[width=0.55\linewidth]{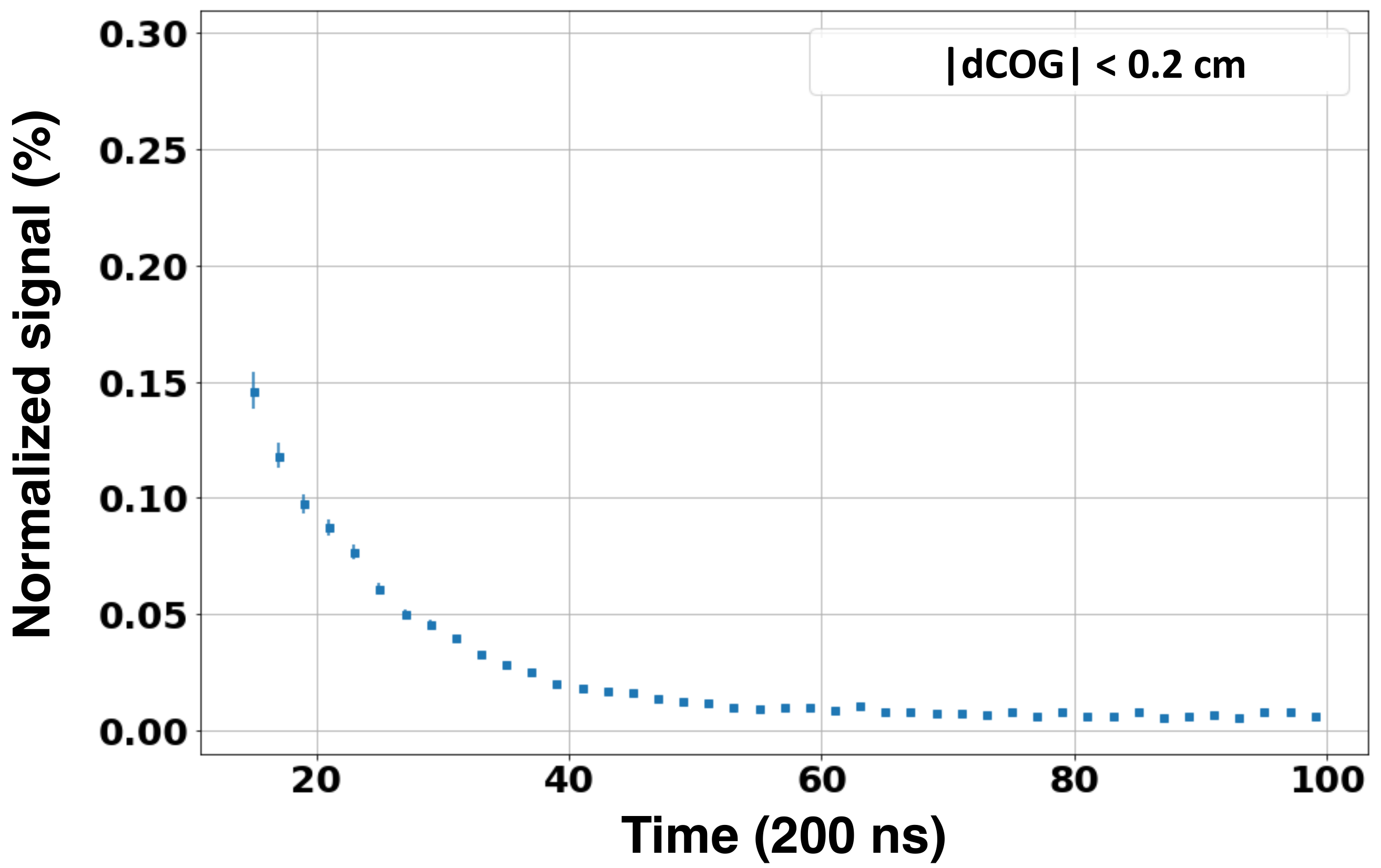}
	\caption{Average normalized ion tail for $E^{50\%}_{\mathrm{ind}}$, which is assumed to describe the fast component independent of the induction field setting. Only the data from the central pads of OROC3 are shown. The error bars correspond to the RMS of entries in each time bin.}
	\label{fig:Eind50}
\end{figure}

\begin{figure}[h]
	\centering
	\includegraphics[width=\linewidth]{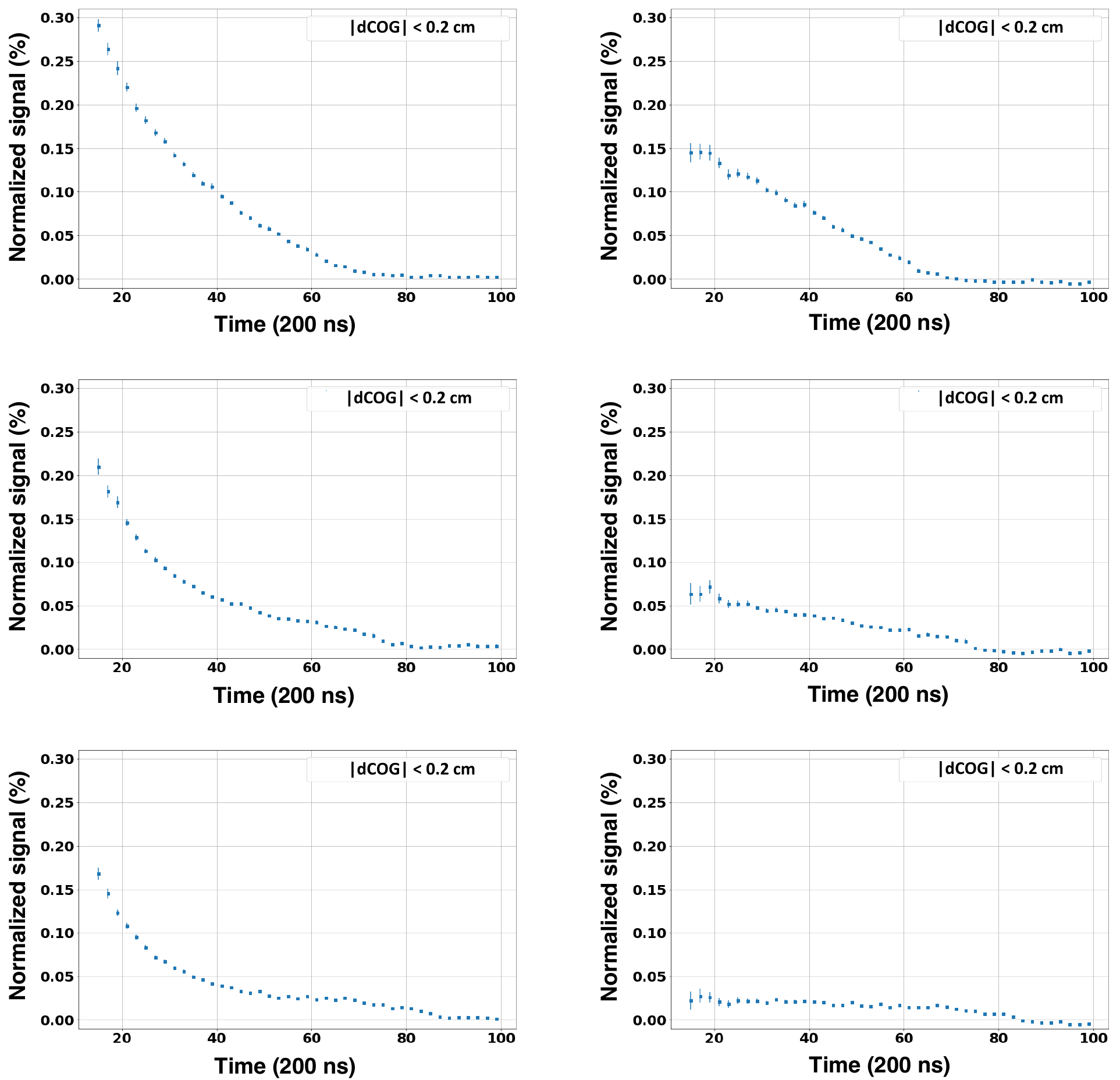}
	  \caption{Average normalized ion tail (left) and difference from the $E^{50\%}_{\mathrm{ind}}$ (right), for $E^{100\%}_{\mathrm{ind}}$ (top), $E^{85\%}_{\mathrm{ind}}$ (middle), $E^{75\%}_{\mathrm{ind}}$ (bottom). Only the data from the central pads of OROC3 are shown. The error bars correspond to the RMS of the entries in each time bin. The slow component of the ion tail (right) is nearly linear for all induction field settings, with its contribution decreasing as $E_{\mathrm{ind}}$ decreases.}
	\label{fig:tails}
\end{figure}

The results of the above procedure are shown for all central pads of OROC3 in \figref{fig:Eind50} and \figref{fig:tails}. In \figref{fig:Eind50}, the average normalized ion tail is shown for the $E^{50\%}_{\mathrm{ind}}$ setting, which is dominated by the fast component. An exponential shape is observed. In \figref{fig:tails}, the normalized ion tail (left) and its difference from the $E^{50\%}_{\mathrm{ind}}$ setting (right) are shown for different values of the induction field. The difference approximates the slow component. A nearly linear shape of the slow component is observed in the right column of~\figref{fig:tails}, consistent with ions uniformly produced in the induction gap. 

\subsection{Ion tail parametrization} \label{itFits}
\begin{figure}[h]
	\centering
		\includegraphics[width=0.6\linewidth]{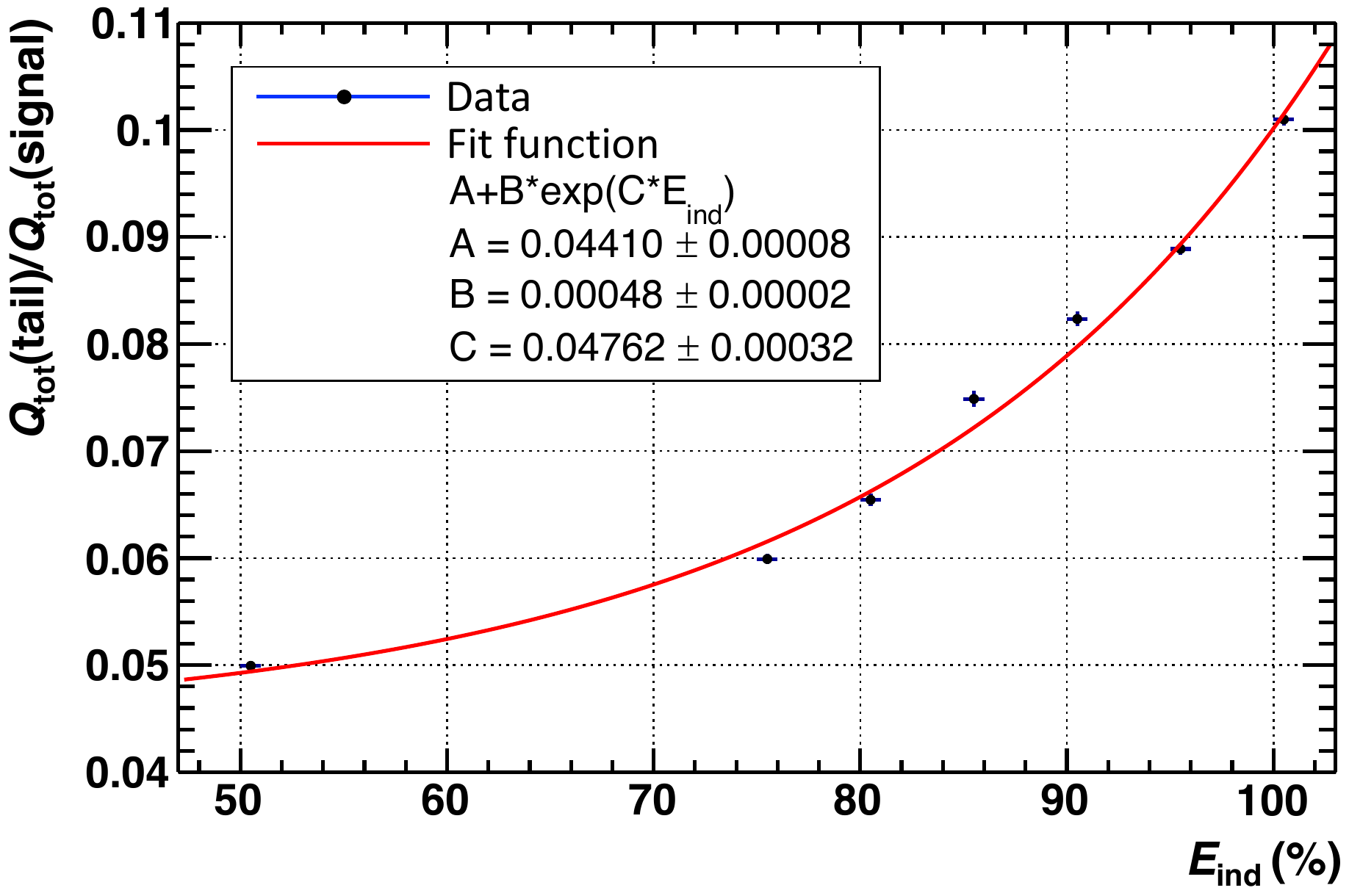}
	\caption{Dependence of the (average) ion-tail fraction on the induction field value. The data points are fit with the function defined in \Eqnref{eq:exp}.}
	\label{fig:ITfractionVsEind}
\end{figure}

The ion tail can be corrected online using an exponential filter algorithm (see \Secref{corrections}). However, the additional linear component must be taken into account to avoid any bias. For this, two external parameters, the \textit{ion-tail slope} and the \textit{ion-tail fraction} ($Q_{\mathrm{tot}}^{\mathrm{tail}}/Q_{\mathrm{tot}}^{\mathrm{signal}}$), are used as input during online processing. To optimize these parameters, each pad signal was fitted with the convolution of a Gaussian and an exponential function to simultaneously describe the signal and tail as shown in the left and right panel of \figref{fig:ionTail_sector_19_row_100_pad63_both}, respectively.
In \figref{fig:ITfractionVsEind}, the ion-tail fraction averaged over all pads is shown as a function of the induction-field setting. The points are fitted with a function of the following form:  
\begin{align}
    f_{\mathrm{IT}}(E_{\mathrm{ind}}) = A + B\cdot e^{C\cdot E_{\mathrm{ind}}} \hspace{0.1cm} .
    \label{eq:exp}
\end{align}
The first term corresponds to the contribution from the GEM4 holes that is independent of the value of the induction field, while the second term corresponds to the contribution from the induction gap. By reducing the induction field value from 100\% to 95\%, the average ion-tail fraction is reduced by 10\%. This indicates that the ion-tail fraction near the nominal setting is very sensitive to small fluctuations in the effective induction field. From the fit, one can also obtain $f_{\mathrm{IT}}(0) \approx 0.045$ and $f_{\mathrm{IT}}(50) \approx 0.05$, indicating that for $E_{\mathrm{ind}}< 50\%$ there is a residual contribution of about 10\% from the slow component. This negligible residual is consistent with the assumption used in \Eqnref{eq:components}. Moreover, the data shown here are consistent with the observation of \figref{fig:tails}, where for the nominal induction field the contributions of the fast and slow components are almost equal, $f_{\mathrm{IT}}(50)/f_{\mathrm{IT}}(100) \approx 0.5$.

\begin{figure}[h]
	\centering
	\includegraphics[width=\linewidth]{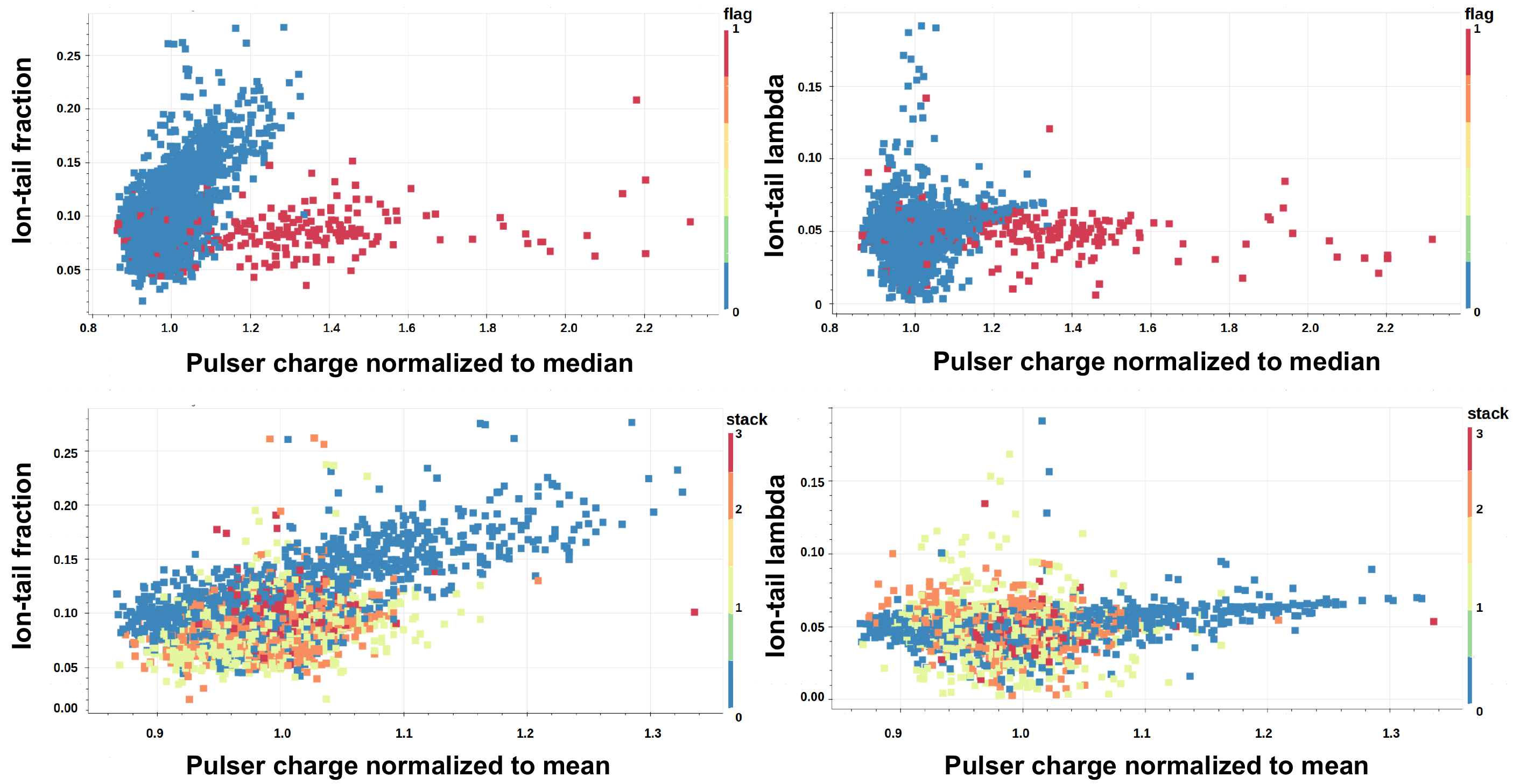}
    \vspace{-0.5cm}
	\caption{(Color online) Ion-tail fraction (left) and ion-tail slope (right) as a function of $Q^{\mathrm{norm}}_{\mathrm{pulser, pad}}$. In the top panels, the color indicates the pad position in the chamber, with blue representing the bulk and red the regions influenced by the presence of the spacer cross or edges. In the bottom panels, the bulk region for pads with $|\mathrm{dCOG}|\leq$ 0.8~cm are shown, where the color scale indicates the stack number.}
	\label{fig:tailParams_dependencies}
\end{figure}

\Figref{fig:tailParams_dependencies} shows the dependence of the ion-tail fraction and the ion-tail slope on the normalized pulser charge. The pads located in edge regions or close to the spacer cross of each chamber were considered separately in the analysis, since the dielectric material placed in these regions influences the signal shapes. In the upper panels of the Figure, these pads (\textit{cross/edge}) are shown in red, while the rest of the pads (\textit{bulk}) are shown in blue. A linear correlation is observed between the ion-tail fraction and the normalized pulser charge for the bulk pads. This is explained by the fact that the pulser charge and the value of the effective induction field are inversely proportional to the distance between the pad and GEM4B, resulting in a wide range of the ion-tail fraction, from about 5 to 20\%. The proportionality does not apply to the pads in the cross/edge region due to the additional dielectric material. In the bottom panels, the bulk region is shown for pads with a distance to the COG of the cluster (dCOG) of less than 0.8~cm and for the four different GEM stacks.

\begin{figure}[h]
	\centering
		\includegraphics[width=0.75\linewidth]{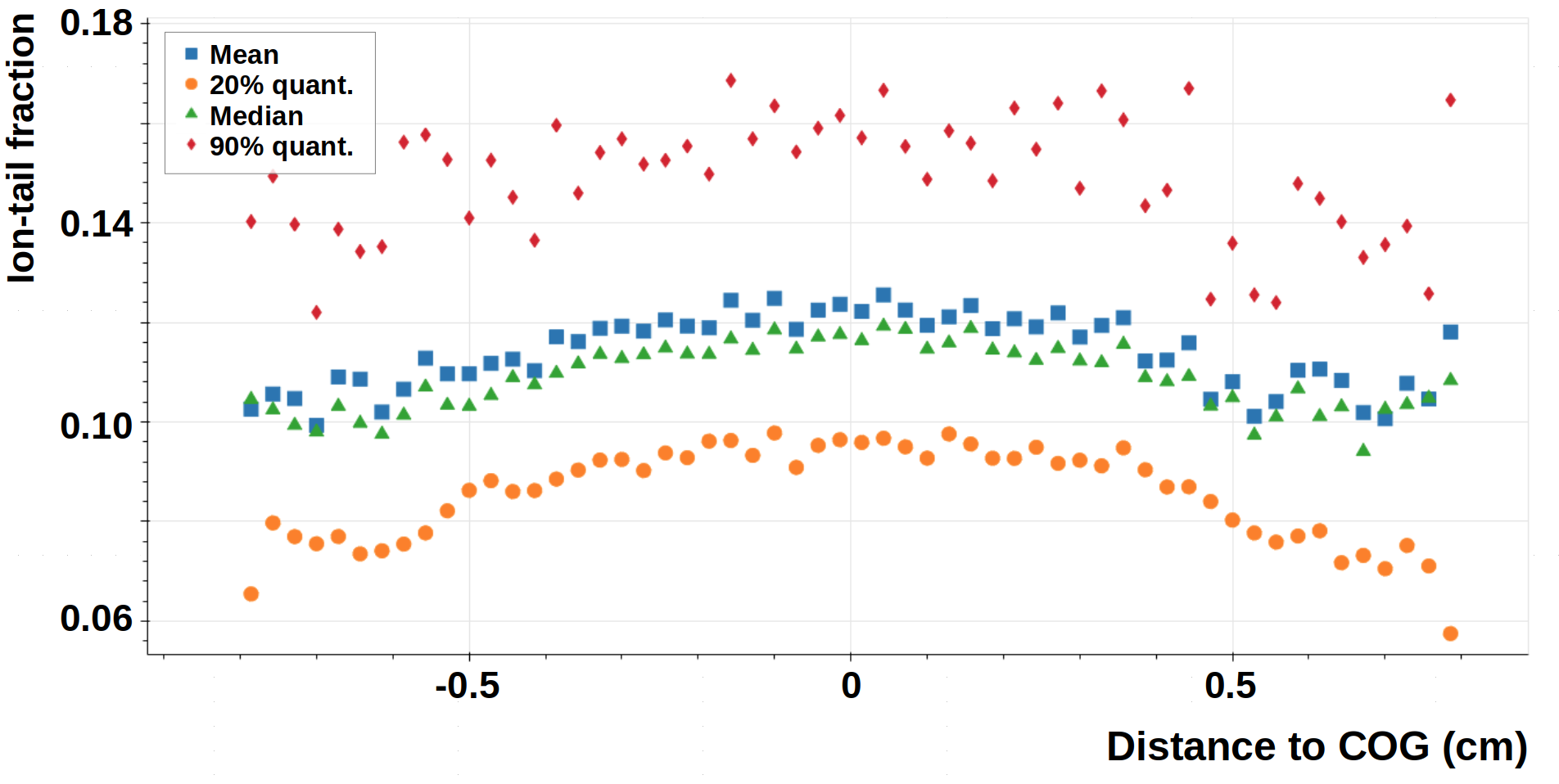}
	\caption{(Color online) Dependence of the (average) ion-tail fraction on the distance of the pad to the COG of the cluster for the bulk pads. Only pads with $|\mathrm{dCOG}|\leq$~0.8~cm are shown. The different quantiles are shown with different colors.}
	\label{fig:deltaCOG} 
\end{figure}

\Figref{fig:deltaCOG} shows the dependence of the ion-tail fraction on the distance from the COG of the cluster for the bulk pads. For the mean, about a $20\%$ difference is observed between the center of the cluster and the center of the neighboring pad. The large spread of the two distributions shown in \figref{fig:tailParams_dependencies}, mainly due to the pad-to-GEM4B distance, does not allow an accurate pad-by-pad calibration of the ion-tail parameters. This is because the foil sagging not only affects the capacitive coupling, but also increases the effective induction field, which leads to a stronger amplification and thus to a change in the ion-tail shape. Moreover, only a small fraction of about 10\% of the TPC pads ``see" the laser signals. Since accurate knowledge of the parameters is important for the restoration of the baseline bias and its fluctuations, krypton calibration data were used to disentangle the ion-tail dependencies (see \Secref{toyMC}).

\section{Common-mode and ion-tail corrections} \label{corrections}
Correcting the common-mode effect and ion tail online, before applying the ZS, is critical for maintaining the PID and tracking performance of the TPC and for limiting cluster losses. Moreover, the correction of the ion tail also helps with minimizing the data volume produced by the TPC. The multiplexed data streamed from the FEE are decoded in the CRUs, where subsequently the pedestal subtraction, common-mode correction, ion-tail correction, and ZS are performed. Due to the large number of pads, the data from each TPC stack are read out by either two (for OROC) or four (for IROC) CRUs.

With the current configuration, information cannot be exchanged between different CRUs. This additional CRU segmentation implies that calculation of the common-mode charge for a pad using \Eqnref{eq:cf0} cannot be applied. The calculation of the average positive signal in the stack $\left<Q_{\mathrm{pos}}(t)\right>_{\mathrm{stack}}$ would require combining information from different CRUs. Instead, a baseline estimation is performed using the \textit{empty} (or \textit{non-signal}) pads in the CRU, by using \Eqnref{eq:cf0} and \Eqnref{eq:cf1}~\cite{Appelshaeuser:2231785}:
\begin{align}
Q_{\mathrm{pad}}^{\mathrm{CM}}(t) = Q^{\mathrm{norm}}_{\mathrm{pulser, pad}}  \left<\frac{Q_{\mathrm{pad}}(t)}{Q^{\mathrm{norm}}_{\mathrm{pulser, pad}}} \right>,
\label{eq:cf3}
\end{align}
where $\left<\right>$ denotes averaging over empty pads in a given CRU. The algorithm consists of two parts: First, the empty pads are selected and the mean baseline is calculated for a given time bin. Second, the common-mode correction is applied to all pads, scaling accordingly using the normalized pulser charge. Since each pad has a different pulser charge, a static map must be provided to the CRUs. An advantage of this method is that the stack-dependent parameters (see \figref{fig:cf_Qpulser}) are not needed, since the correction is calculated at the CRU level. The efficiency of the algorithm depends on the correct selection of the empty pads on a time-bin basis. For this, apart from a simple threshold cut ($Q_{\mathrm{pad}}(t) \leq n \cdot \left<\mathrm{noise} \right>$), an additional check is performed by comparing the pad charge to that of a number of randomly selected pads in the CRU. This ensures that pads measuring the ion tail of earlier signals (superimposed with common-mode) are excluded. A pseudo-code for the common-mode correction is given in~Appendix~\ref{CMCorrCode}.

Online correction of the ion tail is performed on a pad-by-pad basis prior to ZS in the CRUs. Since the polarity of the ion tail is positive, ZS does not result in missing charge and thus missing clusters. Consequently, the ion tail can be corrected online to first order and the remaining second-order deficiencies can be handled during track reconstruction, if necessary. An exponential correction of the form
\begin{align}
Q_{\mathrm{out}}(t) = Q_{\mathrm{in}}(t) - \sum_i A_i\cdot e^{-\lambda_i(t-t_{\mathrm{i,max}})}\hspace{0.2cm}
\label{eq:expCor}
\end{align}
is quite difficult to implement directly in the CRU FPGAs. In the equation, $Q_{\mathrm{in}}(t)$ and $Q_{\mathrm{out}}(t)$ are the pad signals for the time bin $t$ before and after the correction is applied, respectively. The sum runs over all previous signal peaks $i$ for the given pad. The parameters $A_i$ and $\lambda_i$ are the maximum and slope of the tail corresponding to the signal peak $i$, respectively, while $t_{\mathrm{i,max}}$ is the position of the peak maximum. The complexity of such a correction stems from two factors: the number of resources required and the time needed to perform the calculations. First, applying the above correction would mean that the entire peak history for each of the roughly 1600 pads read out by a CRU would need to be stored. However, FPGAs are not typically designed to store large amounts of data. Second, while it is possible for the FPGAs to compute an exponential (e.g., using the CORDIC\footnote{The \emph{CORDIC} (Coordinate Rotation Digital Computer) algorithm is an iterative method for computing elementary functions in the electronics using rotations.} functions ), FPGAs would need to perform multiple exponential calculations in parallel. Based on the number of digital-signal processors available in the CRU FPGA, these calculations would introduce some latency.

To avoid the aforementioned complications, an exponential filter has been developed for online correction of the ion tail. The filter requires only simple mathematical operations. Note that this correction assumes a perfectly exponential ion tail, which is not entirely realistic (see \Secref{itcontr}).
Moreover, the input parameters are tuned using the laser tracks (see \Secref{itFits}), so different track topologies are not taken into account. Namely, residual biases, expected to be on the per-mill level, are inevitable. However, they may still have an impact on the tracking efficiency and PID performance. More details on the implementation of the exponential filter are given in~Appendix~\ref{ITCorrCode}.

\subsection{Toy MC simulations}
\label{toyMC}
To quantify the effects of common-mode and ion tail on the baseline, and to investigate the performance of the two online correction algorithms, toy MC simulations were performed. The stepwise procedure is as follows:
\begin{figure}[t]
  \centering
  \includegraphics[width=\linewidth]{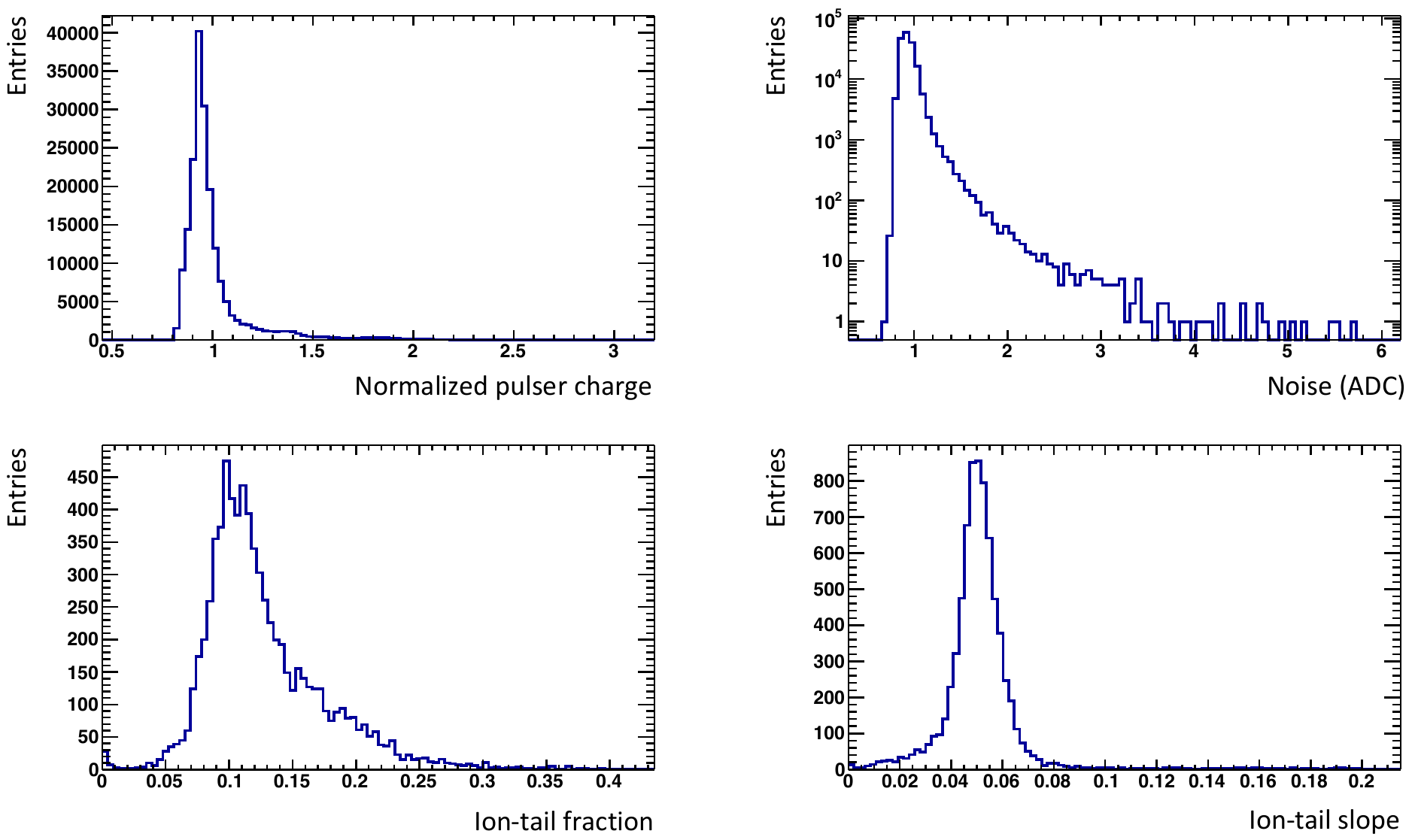}
  \vspace{-0.5cm}
  \caption{Parameter distributions used as input for the toy MC simulation. Only IROC pads are included. Top left: distribution of normalized pulser charge from a pulser calibration run. Top right: noise distribution from a pedestal/noise run. Bottom left (right): ion-tail fraction (slope) obtained from the laser data. For simplicity, no correlations between parameters were assumed.} 
  \label{fig:padParamsMC}
\end{figure}
\begin{figure}[t]
  \centering
  \includegraphics[width=0.5\linewidth]{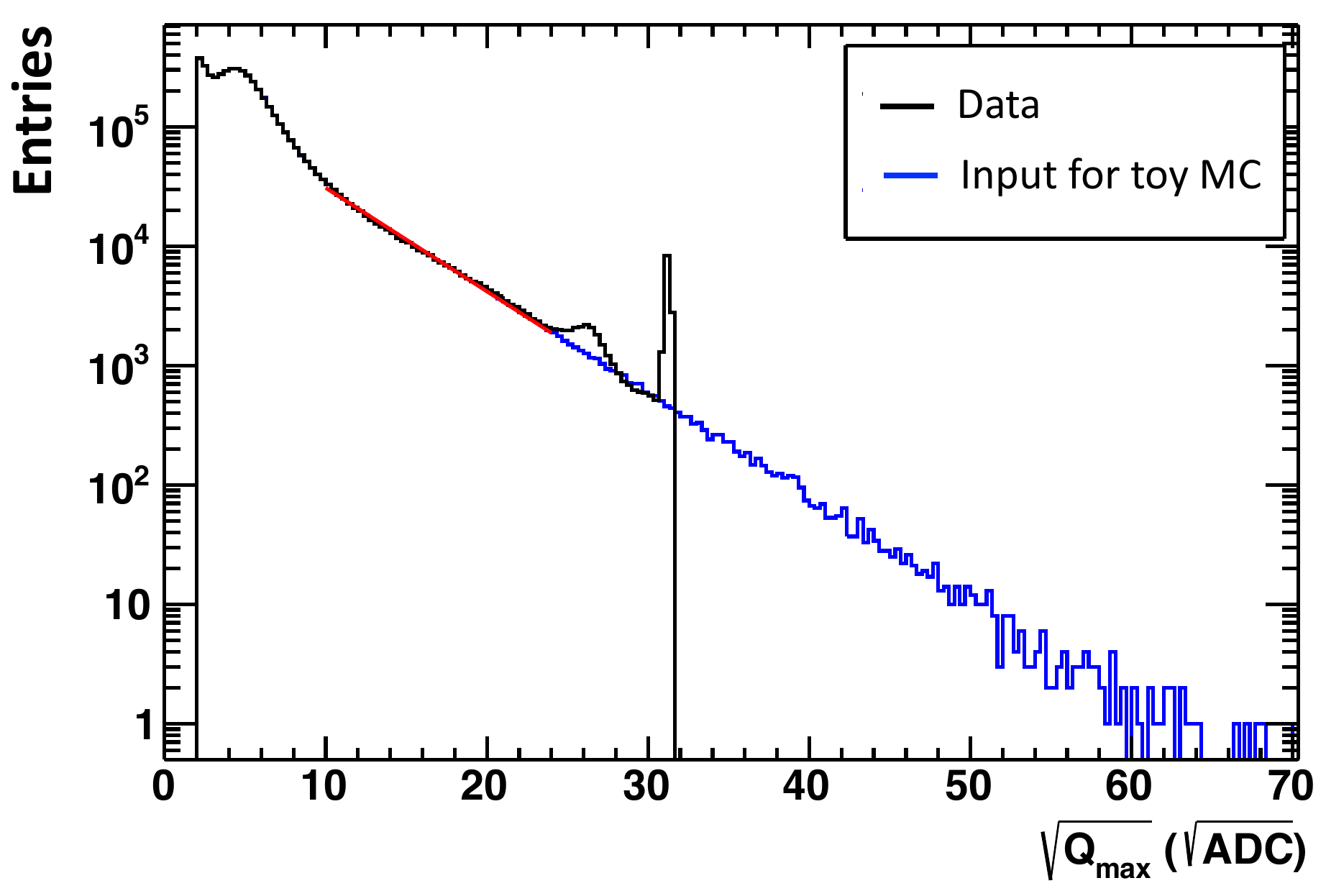}
  \caption{(Color online) Distribution of $Q_{\mathrm{max}}$ as obtained from a data sample of Pb--Pb collisions (black). The two peaks around $26$ and $31$ are due to saturated signals. The distribution was fitted with an exponential function (red) in the range (10, 24). The input distribution for the MC (blue) is constructed using the data distribution for $\sqrt{Q_{\mathrm{max}}}<20$, and the fit function for $\sqrt{Q_{\mathrm{max}}}>20$.} 
  \label{fig:QmaxData}
\end{figure}
\begin{itemize}[label=\textbullet]
    \item 250 Pb--Pb events were simulated, with particle composition derived from particle spectra measured with the ALICE detector at 5.02~TeV (Run~2 conditions). An area of 1000 pads (data from between 1200 and 1600 pads are transferred to a single CRU) and 1000 time bins were considered for the simulation of the common-mode effect.
    \item Random values for the normalized pulser charge, noise, ion-tail fraction and slope were assigned to each pad according to the distributions obtained from the calibration data (see \figref{fig:padParamsMC}). For the pedestal, a value in the range (0,1) was assigned, since the integer part of the pedestal can be subtracted trivially.
    \item A number of clusters were generated, with the number of clusters per event randomly distributed such that the variations in detector occupancy resembled the conditions of Run~3.
    \item For each cluster, a value for $Q_{\mathrm{max}}$ (maximum value among all digits in a cluster) was generated using the distribution obtained from the Pb--Pb collision data (see \figref{fig:QmaxData}). Each cluster spreads over three pads and three time bins, following a Gaussian distribution in the pad- and time-space.
    \item The ion tail was simulated as a perfect exponential, and for each pad, the simulated parameters for the ion-tail slope and fraction were used.
    \item The common-mode was simulated using \Eqnref{eq:cf1}. A value of 0.5 was used for the slope, namely $k_{\mathrm{CF, pad}} = 0.5 \cdot Q^{\mathrm{norm}}_{\mathrm{pulser, pad}}$.
    \item Noise, pedestal and rounding were added.
    \item The pedestals were subtracted. 
    \item The common-mode effect was corrected using the algorithm described in~Appendix~\ref{CMCorrCode}. 
    \item The ion tail was corrected using the algorithm described in~Appendix~\ref{ITCorrCode}. The order in which the ion-tail and common-mode corrections are applied is important because the ion tail also produces small common-mode undershoots in the other pads.
\end{itemize}
\begin{figure}[h]
	\centering
	\includegraphics[width=0.65\linewidth]{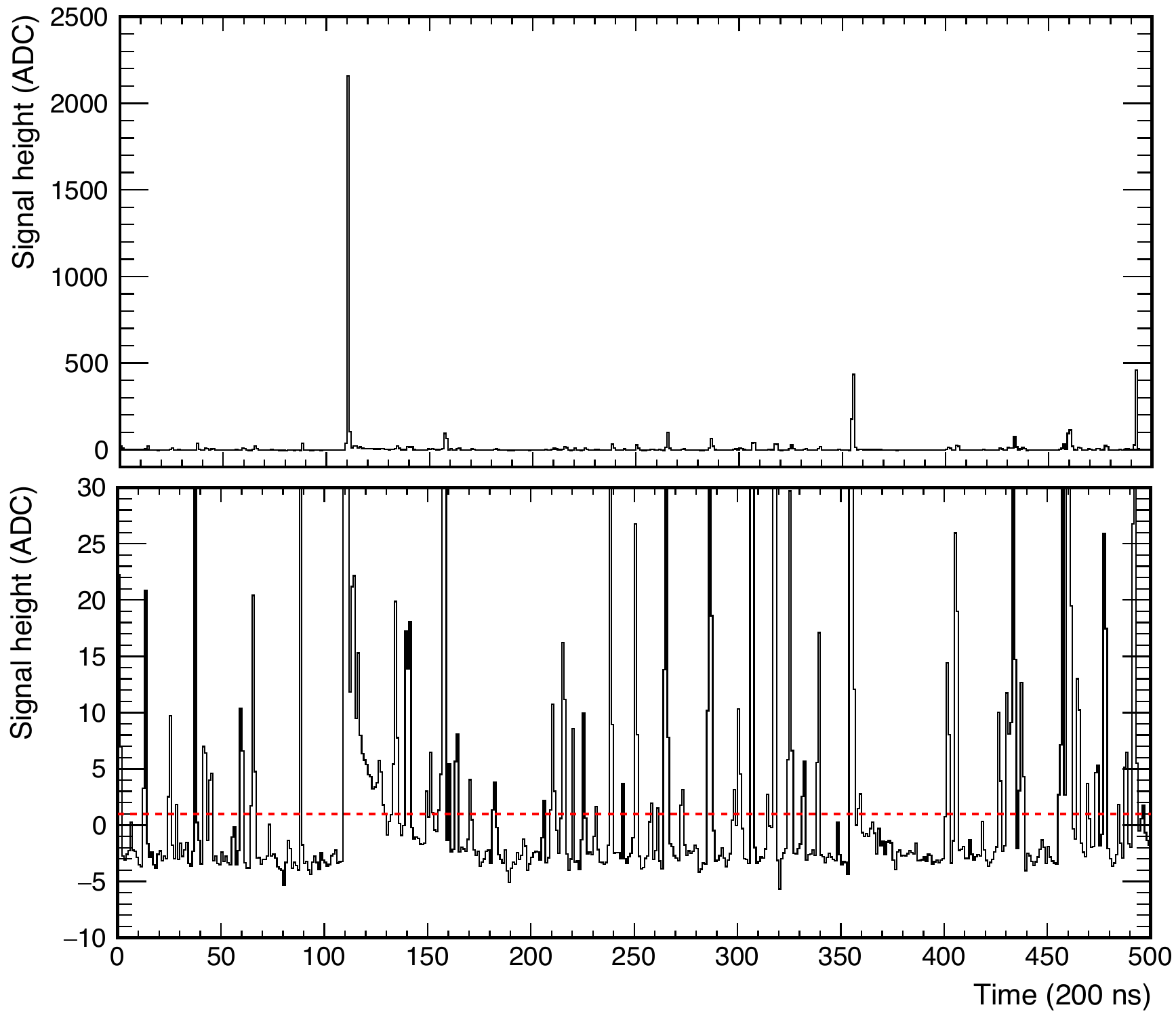}
	\caption{Top: pad signal for an event with about 30\% occupancy, without noise. Bottom: pad signal zoomed in on the $y$-axis. The red dashed line shows the ZS threshold.}
	\label{fig:run3_baseline}
\end{figure}

\Figref{fig:run3_baseline} shows a simulated pad signal for an event with multiplicity corresponding to approximately 30\% occupancy in TPC. Both the common-mode effect and the ion tail are included. The baseline is systematically shifted to negative values due to the common-mode effect. Next, a series of simulations were performed to optimize the correction parameters, including simulated noise. For each event, 500 random settings were simulated, allowing for the following options:
\begin{itemize}[label=\textbullet]
    \item Common-mode simulation: ON or OFF.
    \item Common-mode simulation ON and correction of the common-mode effect either OFF, or using a mean correction, median correction, mean $2^{\mathrm{nd}}$ iteration, or median $2^{\mathrm{nd}}$ iteration.
    \item Random values (in reasonable ranges) for the parameters \texttt{nPadsRandom, nPadsMin, Q\_thr1, Q\_thr2}.
    \item Ion-tail simulation: ON or OFF.
    \item Ion-tail simulation ON and ion-tail filter either OFF, or correction using the pad-by-pad ion-tail parameters (\textit{pad-by-pad}), or correction using the median value of the parameter distributions shown in the bottom of \figref{fig:QmaxData} (\textit{fixed-to-median}).
    \item Random value for \texttt{k0} in the range 70--100\%.
\end{itemize}
\begin{figure}[b]
	\centering
	\includegraphics[width=\linewidth]{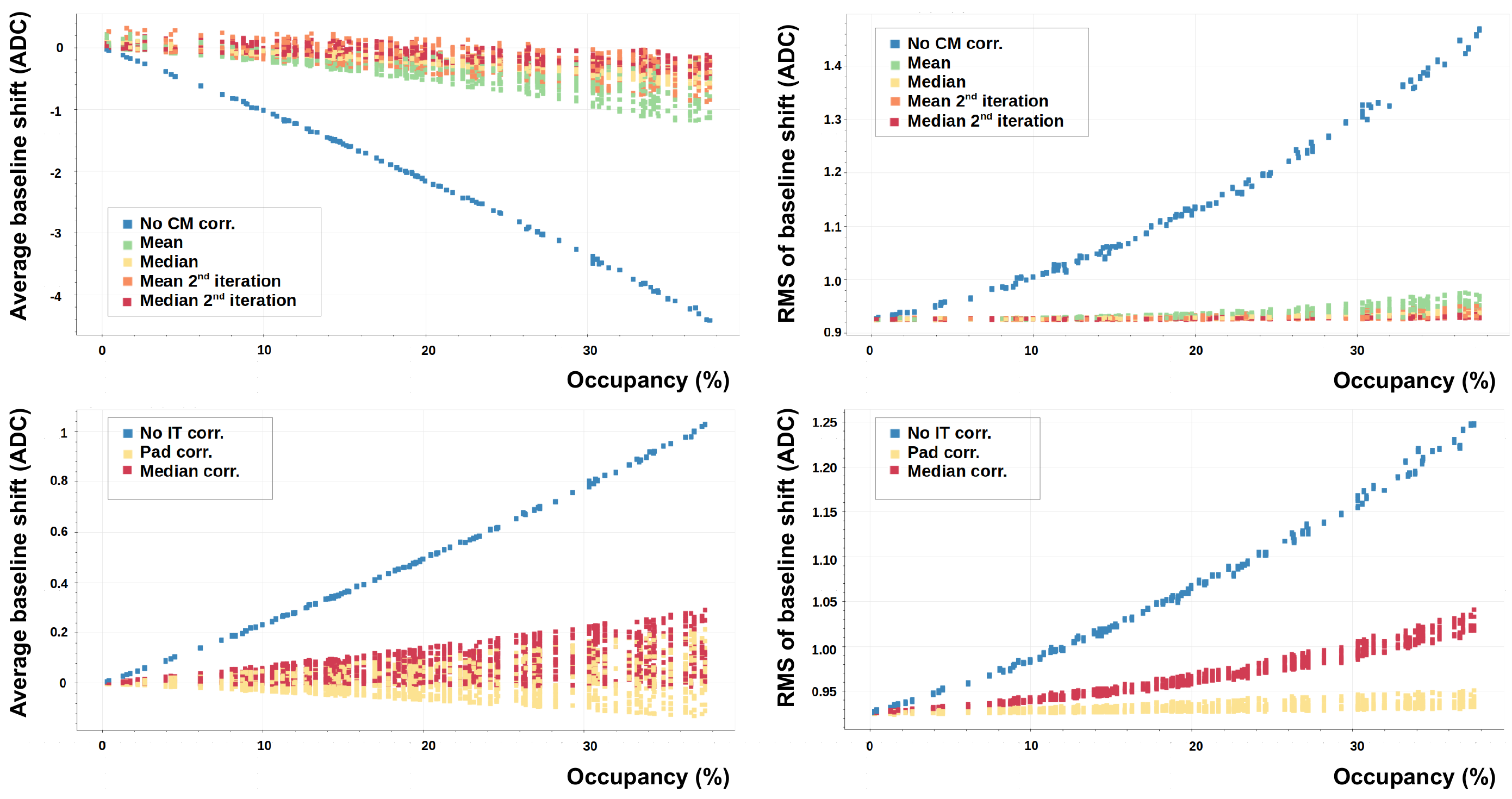}
    \vspace{-0.5cm}
	\caption{(Color online) Average baseline shift (left) and its RMS (right) as a function of occupancy. The simulation and correction of the "common-mode only" (“ion tail only”) scenario is shown in the top (bottom) panel. The various marker colors corresponds to the different correction methods.}
	\label{fig:mcScan1}
\end{figure}
\Figref{fig:mcScan1} shows the results of the parameter scan. On the left, the average baseline shift and on the right its RMS are shown as a function of occupancy. A 99\% least-trimmed-squares method was used to exclude some extreme outliers. For the top panels, only the common-mode effect was simulated. The different correction methods (no correction, mean, median, and $2^{\mathrm{nd}}$ iterations) are shown with different colors. It can be seen that the average baseline bias can reach up to $-5$~ADC and its RMS up to $1.5$~ADC if the common-mode effect is not corrected. Compared to the RMS in the absence of the two effects (caused by the noise), this corresponds to roughly a 60\% increase. For the bottom panels, only the ion tail was simulated. The average baseline bias reaches up to $1$~ADC and its RMS $1.25$~ADC, which corresponds to an increase of about $35\%$ when the tail is not corrected.

\begin{figure}[h!]
	\centering
	\includegraphics[width=\linewidth]{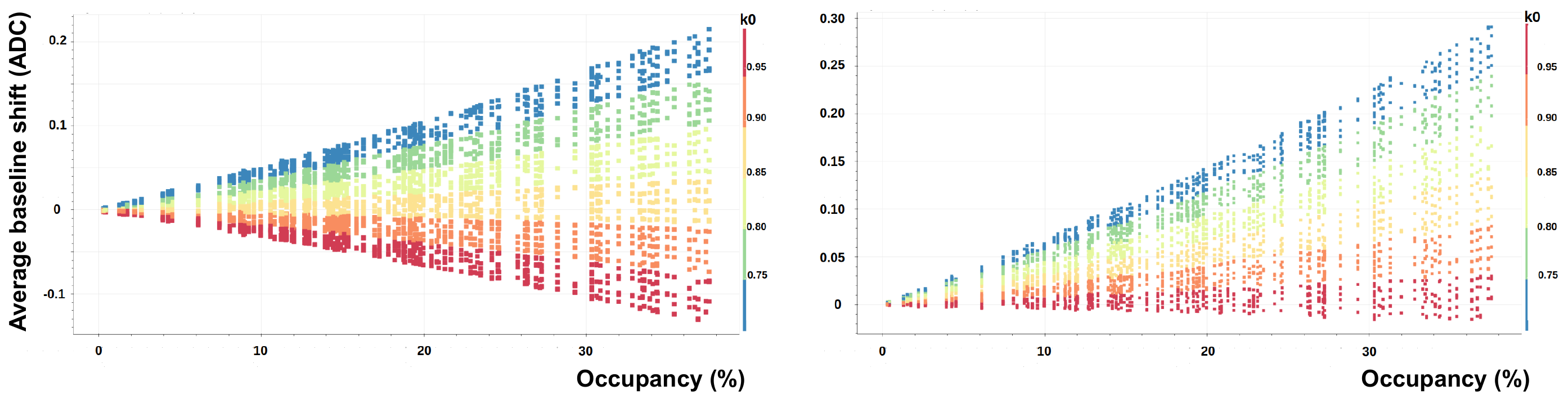}
    \vspace{-0.5cm}
	\caption{(Color online) Average baseline shift as a function of occupancy for the pad-by-pad (left) and the fixed-to-median (right) methods of the ion-tail correction. The common-mode effect was not simulated. The color scale indicates the different correction fractions.}
	\label{fig:mcScan4}
\end{figure}

In \figref{fig:mcScan4}, the average baseline shift is shown as a function of occupancy for the pad-by-pad (left) and the fixed-to-median (right) corrections of the ion tail. As mentioned in~Appendix~\ref{ITCorrCode}, a value of \texttt{k0=100\%} overcorrects the data, which can be confirmed by the negative values of the baseline bias (red markers). It can be seen that the optimal value for the pad-by-pad method is \texttt{k0}~$\approx$~\texttt{85-90\%}, while for the fixed-to-median method it is \texttt{k0}~$\approx$~\texttt{90-95\%}. The baseline shift is slightly larger for the fixed-to-median method.

\begin{figure}[h!]
	\centering
	\includegraphics[width=\linewidth]{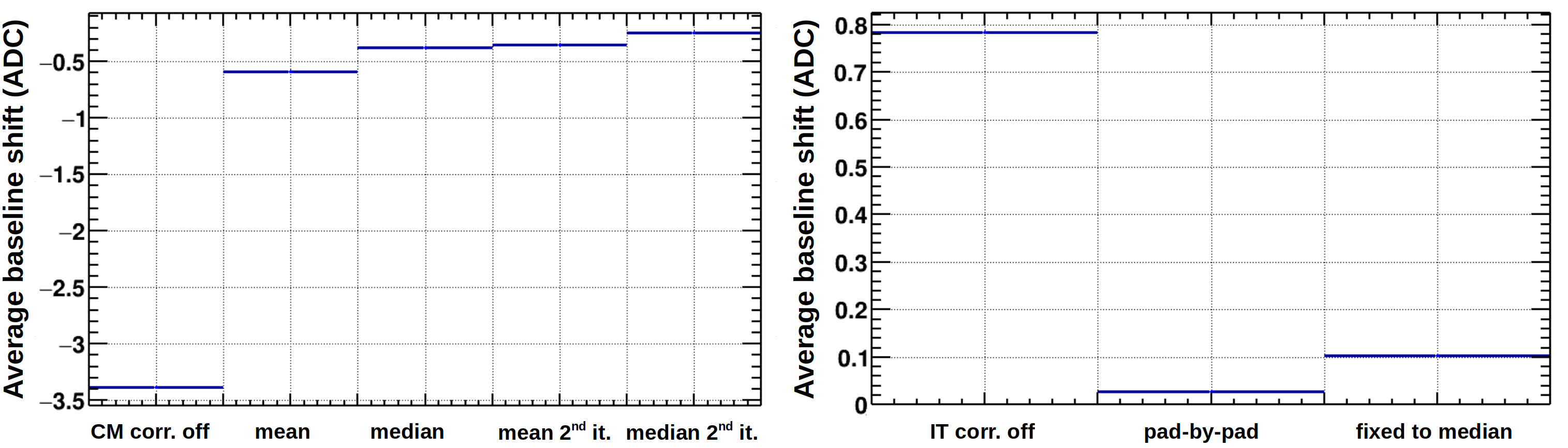}
    \vspace{-0.5cm}
	\caption{Average baseline shift for the different methods used in  the common-mode (left) and ion-tail (right) corrections, where the occupancy is 28--32\%. All parameter settings were averaged.}
	\label{fig:baselineMCProfs}
\end{figure}

\Figref{fig:baselineMCProfs} shows the average baseline shift for the different methods used in the common-mode (left) and ion-tail (right) corrections, where the occupancy is 28--32\%, corresponding to the highest expected multiplicities in Run 3. All parameter settings were averaged. It can be seen that the common-mode correction can restore most of the baseline shift. Using the median $2^{\mathrm{nd}}$ iteration method instead of the mean results in roughly a 60\% smaller average baseline bias. Similarly, correction of the ion tails restores most of the baseline shift. Using the pad-by-pad method instead of the fixed-to-median method results in an approximately 70\% smaller average baseline shift.

\begin{figure}[h!]
	\centering
	\includegraphics[width=0.65\linewidth]{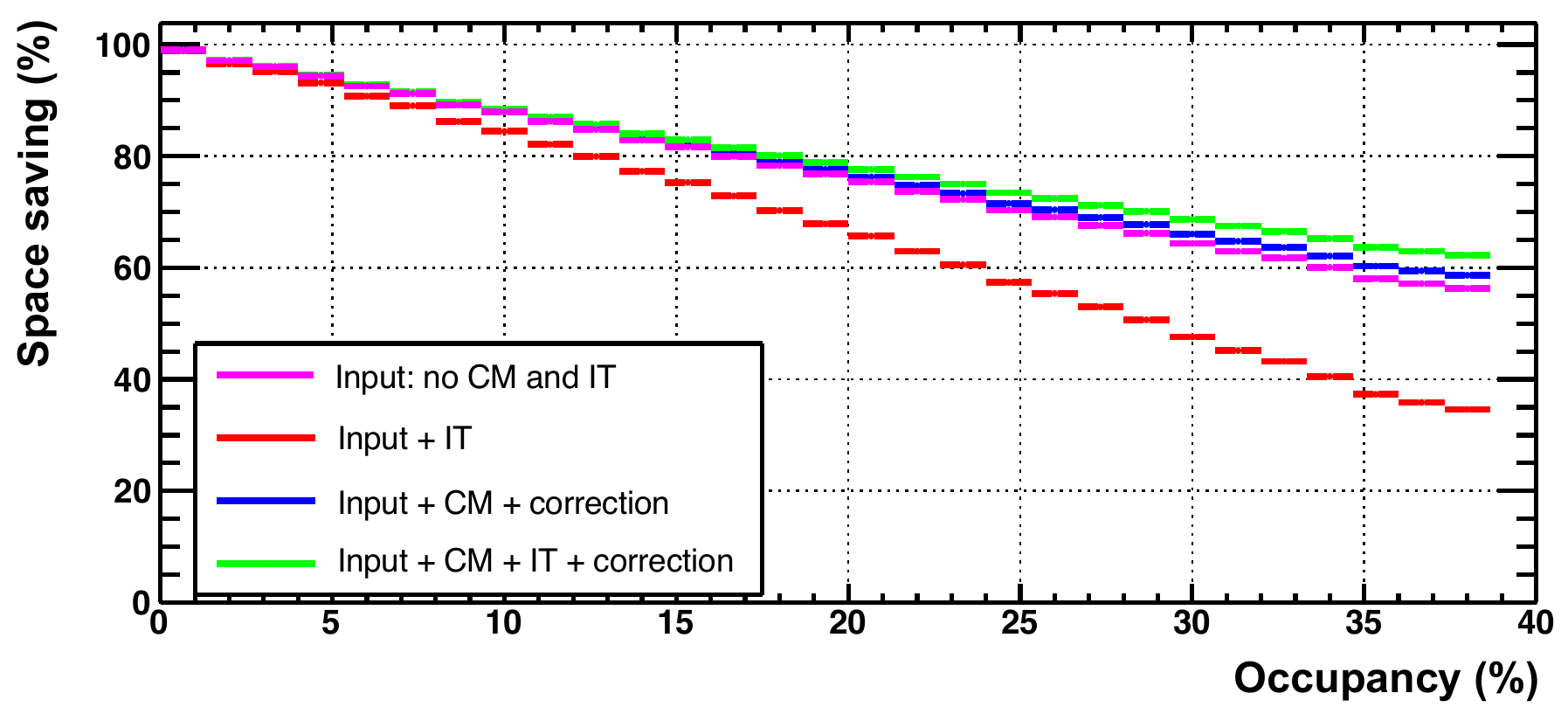}
	\caption{(Color online) Impact of the ion tail on the space saving. Noise was not included, and a threshold cut of 1.2~ADC was applied.}
	\label{fig:impactTail}
\end{figure}

The impact of the ion tail on the data volume produced by the TPC is shown in \figref{fig:impactTail} as a function of occupancy. The \textit{space saving} is defined as:
\begin{align}
    \mathrm{space\ saving} (\%) = 100\% - \frac{\mathrm{compressed\ data\ size}}{\mathrm{uncompressed\ data\ size}} \hspace{0.2cm},
\end{align}
where the uncompressed and compressed data size is the size before and after ZS, respectively. It can be seen that the ion tail significantly reduces the space saving compared to the original signal. As shown by the blue and magenta lines, simulating and correcting for the common-mode effect leads to a larger space saving than the original signal due to a remaining negative bias. The reduction in space saving due to the ion tail can be restored by applying its correction.

\Tabref{tab:impactTail} summarizes the importance of the ion-tail correction for space saving for minimum bias Pb--Pb events (occupancy $\leq 30\%$). Due to the ZS, an uncompressed data rate of 3.5~TB/s would be reduced to 650~GB/s in the absence of the ion tail and the common-mode. This number is consistent with the expected zero-suppressed rate of about 600~GB/s~\cite{Rohr:2021psv}. When adding the ion tail, the rate increases by about 30\%, to 888 GB/s. After correction for both effects, a given choice of parameters results in an inaccuracy of about 3 to 5\% for each effect.

\begin{table}[h!]
\centering
\begin{tabular}{c|cc}
                              & \begin{tabular}[c]{@{}c@{}}Minimum bias\\ space saving (\%) \end{tabular} & GB/s  \\ \hline
input signal                 & 81.4                                                                      & 650\\
input + ion tail                   & 74.6                                                                    & 888\\
input + common-mode + correction      & 82.1                                                                    & 628 \\
input + ion tail + common-mode + correction & 83.1                                                                     & 594 \\ \hline
current assumption& 82.9 & 600 \\ 
\end{tabular}
\caption{Impact of the ion-tail correction on the  space saving for minimum-bias events with occupancy $\leq$30\%. Noise was not included, and a threshold cut of 1.2~ADC was applied. Note the realistic order of implementation of the effects and the corresponding corrections.}
\label{tab:impactTail}
\end{table} 

Although this toy MC simulation study is based on a data-driven approach, detector performance studies using the full MC simulations should be carried out in order to account for different track topologies and cluster shapes, diffusion, and other detector effects that have not been considered.

\section{Conclusions}

The signal response of the GEM-based ALICE TPC was studied in detail using the data collected with the laser calibration system. The dependencies of the common-mode effect were understood using machine learning techniques. It was found that the common-mode signal depends largely (96\%) on the stack type and the normalized charge detected in runs with the dedicated calibration pulser system. The stack type accounts for the absolute capacitance of the stack, while the normalized pulser charge is responsible for the pad-by-pad capacitance variations. An unpredicted ion tail was observed in data recorded with the TPC laser system. The measurements have shown that ions from two categories contribute almost equally to the generated signal at the nominal induction field setting. The contribution of the ions generated in the GEM4 holes is practically independent of the value of the induction field and results in an exponentially shaped tail. Ions generated in the induction gap of the GEM stack result in an additional contribution, which depends on the electric field applied to the induction gap.

The performance of the two online correction algorithms was studied in detail using a toy MC with input parameters determined in a data-driven way. The common-mode correction algorithm correlates all pads of a given CRU for each time bin, while the ion tail is corrected on a per-channel basis using an exponential filter. Both effects are efficiently corrected, despite a residual bias in the baseline, comparable to the noise. Since the ion-tail parameters used in the exponential filter are obtained from the laser tracks, different track topologies were not considered. Therefore, possible imperfections in the ion-tail correction are to be expected. These should be considered when repeating these studies using a full MC simulation. Furthermore, the ion tail has a significant impact on online data compression. During Run~3 and Run~4 data taking periods, the raw data readout rates are estimated to be about 3.5~TB/s. The toy MC results obtained in this study show that in the presence of the ion tail, the final data rate after baseline subtraction is about 890~GB/s instead of 650~GB/s, which is the estimated value given in the technical design report~\cite{TDR:tpcUpgrade}. The proposed ion-tail correction algorithm fully covers this increased data rate. The two correction algorithms will be commissioned using pp and Pb--Pb collisions at record energy and luminosities as part of the TPC readout system from 2023 onward.

\section*{Acknowledgements}
The ALICE TPC Collaboration acknowledges the following funding agencies for their support in the TPC Upgrade:
Funda\c{c}\~ao de Amparo \`a Pesquisa do Estado de S\~ao Paulo (FAPESP), Brasil;
Ministry of Science and Education, Croatia;
The Danish Council for Independent Research | Natural Sciences, the Carlsberg Foundation and Danish National Research Foundation (DNRF), Denmark;
Helsinki Institute of Physics (HIP) and Academy of Finland, Finland;
Bundesministerium f\"{u}r Bildung, Wissenschaft, Forschung und Technologie (BMBF), 
GSI Helmholtzzentrum f\"{u}r Schwerionenforschung GmbH, 
DFG Cluster of Excellence "Origin and Structure of the Universe", 
The Helmholtz International Center for FAIR (HIC for FAIR)
and the ExtreMe Matter Institute EMMI at the GSI Helmholtzzentrum f\"{u}r Schwerionenforschung, Germany;
National Research, Development and Innovation Office, Hungary;
Nagasaki Institute of Applied Science (IIST)
and the University of Tokyo, Japan;
Fondo de Cooperaci\'{o}n Internacional en Ciencia y Technolog\'{i} a (FONCICYT), Mexico;
The Research Council of Norway, Norway; Ministry of Science and Higher Education and National Science Centre, Poland;
Ministry of Education and Scientific Research, Institute of Atomic Physics and Ministry of Research and Innovation, and Institute of Atomic Physics, Romania;
Ministry of Education, Science, Research and Sport of the Slovak Republic, Slovakia;
Swedish Research Council (VR), Sweden;
United States Department of Energy, Office of Nuclear Physics (DOE NP), United States of America. 

\clearpage
\begin{appendices} 
\section{Common-mode correction algorithm} \label{CMCorrCode}
The following pseudo-code for the common-mode correction is tested in \Secref{toyMC} using a toy MC simulation and implemented in the CRU firmware:

\begin{lstlisting}[language=C++]
// Some constants to be set before 
int nPadsCRU; // Number of pads in the current CRU
int nPadsRandom; // Number of pads (around 10) at random distance to the current to pad. To be compared for additional check if the current pad is an empty pad 
int nPadsMin; // Minimum number of pads (around nPadsRandom/2) required to have Q very close to current pad
float Q_thr1; // Optimized threshold 1 (comparable to 2*noise)
float Q_thr2; // Optimized threshold 2 (comparable to noise)
//
// Main CM correction algorithm
for each time bin{
    // Calculate mean baseline of empty pads
    vector<float> Q_pad_array; // Array to hold Q of empty pads of a CRU
    for each pad{
        float k_pulser = GetNormQPulser(padID); // from 2D pad-by-pad map
        // Simple check if the current pad is empty
        if (Q_pad <= Q_thr1){
            // Additional check if the current pad is empty pad 
            int nPadsOK = 0;
            for (int i = 0; i < nPadsRandom; i++){
                int padRandomID = GetRandomPad(); // Randomly select a pad
                float k_pulser_rndm = GetNormQPulser(padRandomID);
                if (abs(Q_pad/k_pulser - Q_padRndm/k_pulser_rndm) < Q_thr2) 
                nPadsOK++; 
            } 
            // If empty, add charge of current pad into array, scale accordingly 
            if (nPadsOK >= nPadsMin) Q_pad_array.push_back(Q_pad/k_pulser);
        } 
    } 
    // Calculate mean baseline of the non-signal pads
    Q_baseline = mean(Q_pad_array); // Alternatively use median
    // Apply common-mode correction
    for each pad{
        float k_pulser = GetNormQPulser(padID); // From 2D pad-by-pad map
        Q_pad = Q_pad - Q_baseline*k_pulser;
    } 
} 
\end{lstlisting}

\begin{figure}[h!]
	\centering
	\includegraphics[width=0.65\linewidth]{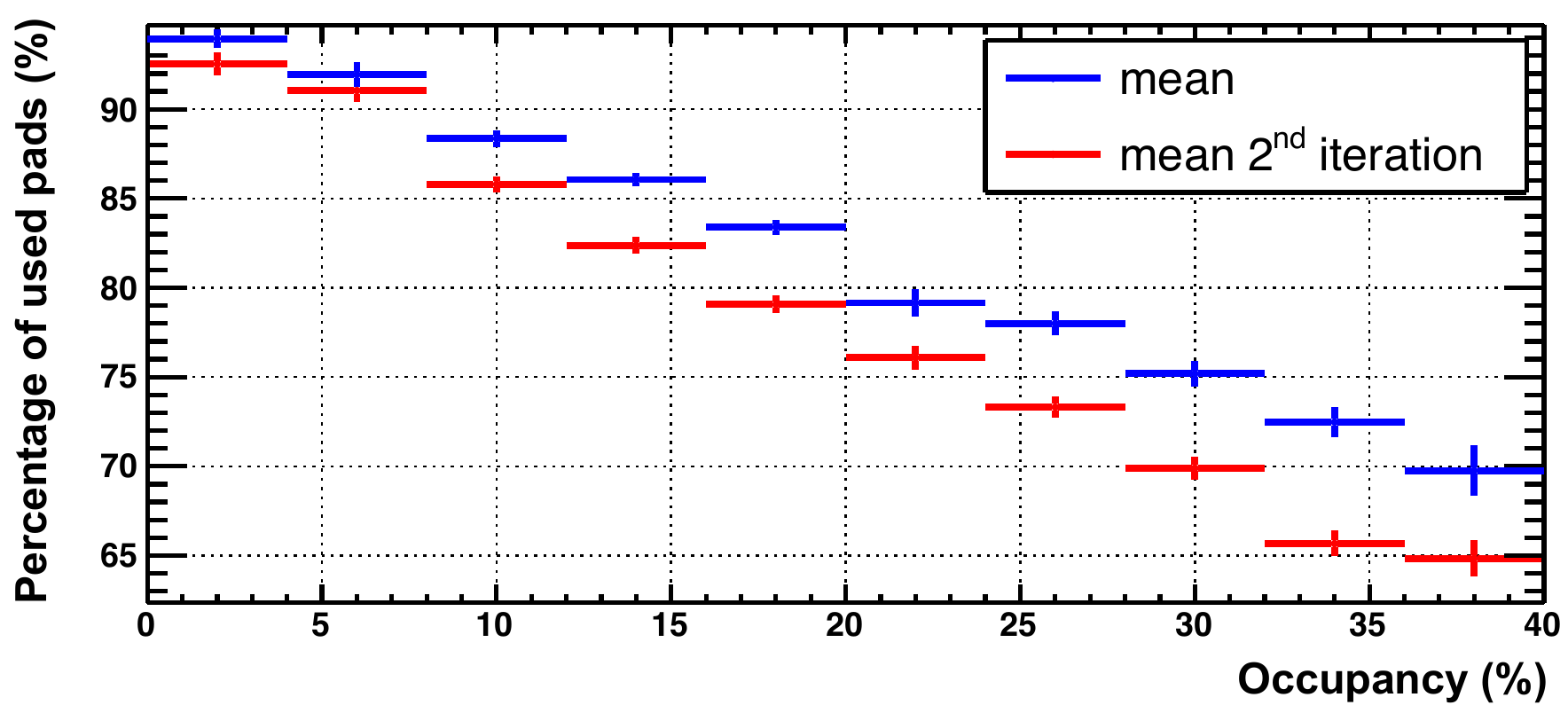}
	\caption{(Color online) Fraction of pads used for the common-mode baseline estimation as a function of occupancy, for the mean (blue) and the mean $2^{\mathrm{nd}}$ iteration (red) methods. Only the common-mode effect is simulated.}
	\label{fig:nOK}
\end{figure}

Note that \texttt{Q\_pad} and \texttt{Q\_padRndm} are the charge after the pedestal subtraction. The above algorithm can be made more robust by using a median estimator instead of the mean for the baseline estimation. Furthermore, since the mean method can incorrectly classify pads with low-signal amplitudes as empty, biasing the baseline estimate, the selection of empty pads can be improved by performing a second iteration after the baseline calculation and before the common mode correction (line 29). In this case, the simple check in line 15 becomes (\texttt{Q\_pad}~-~\texttt{Q\_baseline*k\_pulser}~<=~\texttt{Q\_thr1}), which allows a better classification of the empty pads.  \Figref{fig:nOK} shows the fraction of pads identified as empty pads for the mean and the mean $2^{\mathrm{nd}}$ iteration common-mode correction methods as a function of occupancy. 

\section{Ion-tail correction algorithm} \label{ITCorrCode}
The following pseudo-code for the ion-tail correction is tested in \Secref{toyMC} using a toy MC simulation and implemented in the CRU firmware:

\begin{lstlisting}[language=C++]
// Constant to be set before
float k0; // Multiplicative correction factor, same for all pads (around 0.9)
for each padID{
    float k1 = GetFractionIT(padID); //k_fraction, from a static map
    k1*=k0; // Scale IT fraction
    float k2 = GetExpSlopeIT(padID); // = exp(-k_slope), from a static map
    float Q_correction = 0;
    for each time bin{
        Q_out = Q_in - k1*(1-k2)*Q_correction;
        Q_correction+=Q_in;
        Q_correction*=k2;
    }
}
\end{lstlisting}

The amount of correction for a given time bin is \texttt{k1*(1-k2)*Q\_correction}, where \texttt{Q\_correction} is the buffered charge value of the previous time bin and \texttt{k1} (fraction of the ion-tail integral with respect to the total charge of the pad signal) and \texttt{k2} (slope of the ion-tail assuming an exponential form) are the ion-tail parameters. Since these are pad-dependent parameters, they are provided as input to the CRUs in the form of two-dimensional, pad-by-pad maps. The pad-by-pad calibration of these parameters was obtained from data recorded with radioactive krypton isotopes in the gas volume~\cite{Alme:2010ke}. Alternatively, the median value of the distributions shown in \figref{fig:tailParams_dependencies} can be used. The impact of this simplification is demonstrated in \Secref{toyMC}. An additional scaling parameter, \texttt{k0}, whose value is the same for all pads, is introduced to account for the difference between the integral of the input charge with a continuous Gaussian form and the digitized signal, which has a discrete structure. The digitized signal is equal to or greater than the input, so the parameter \texttt{k0} is defined as \texttt{k0}~$\le1$. Note that the above filter can be applied continuously, even in the absence of signals. In this case, the filter has no effect on the measured charge.

\begin{figure}[h!]
	\centering
	\includegraphics[width=0.7\linewidth]{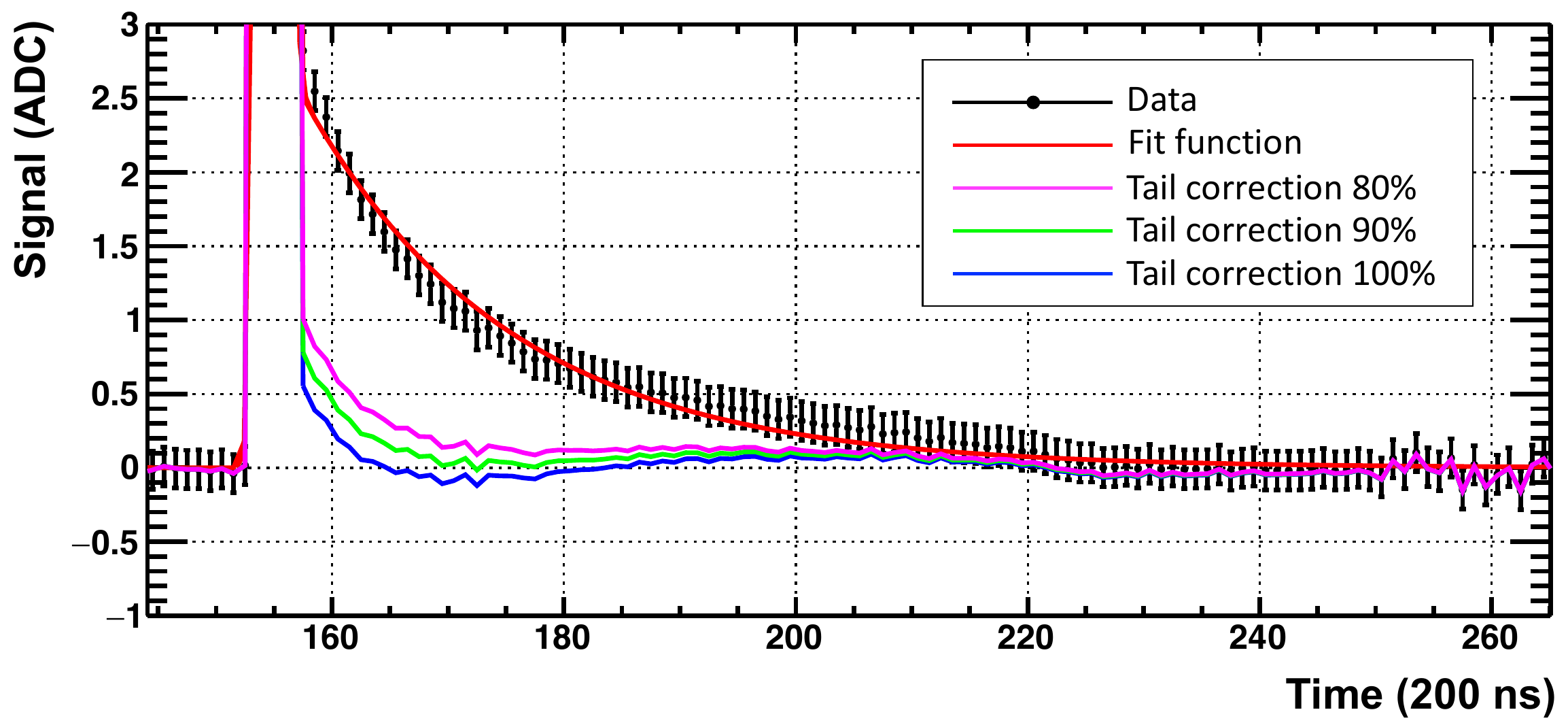}
	\caption{(Color online) Demonstration of the ion-tail filter applied to laser data averaged over many signals on a given pad. The laser data are shown in black, the fit function in red, and the corrected data in blue, green, and magenta, for different values of the parameter \texttt{k0}. The convolution of a Gaussian and an exponential function was used to fit the data. The ion-tail parameters obtained from the fit were used for the correction.}
	\label{fig:tailCorFract}
\end{figure}

\begin{figure}[h!]
  \centering
	\includegraphics[width=0.7\linewidth]{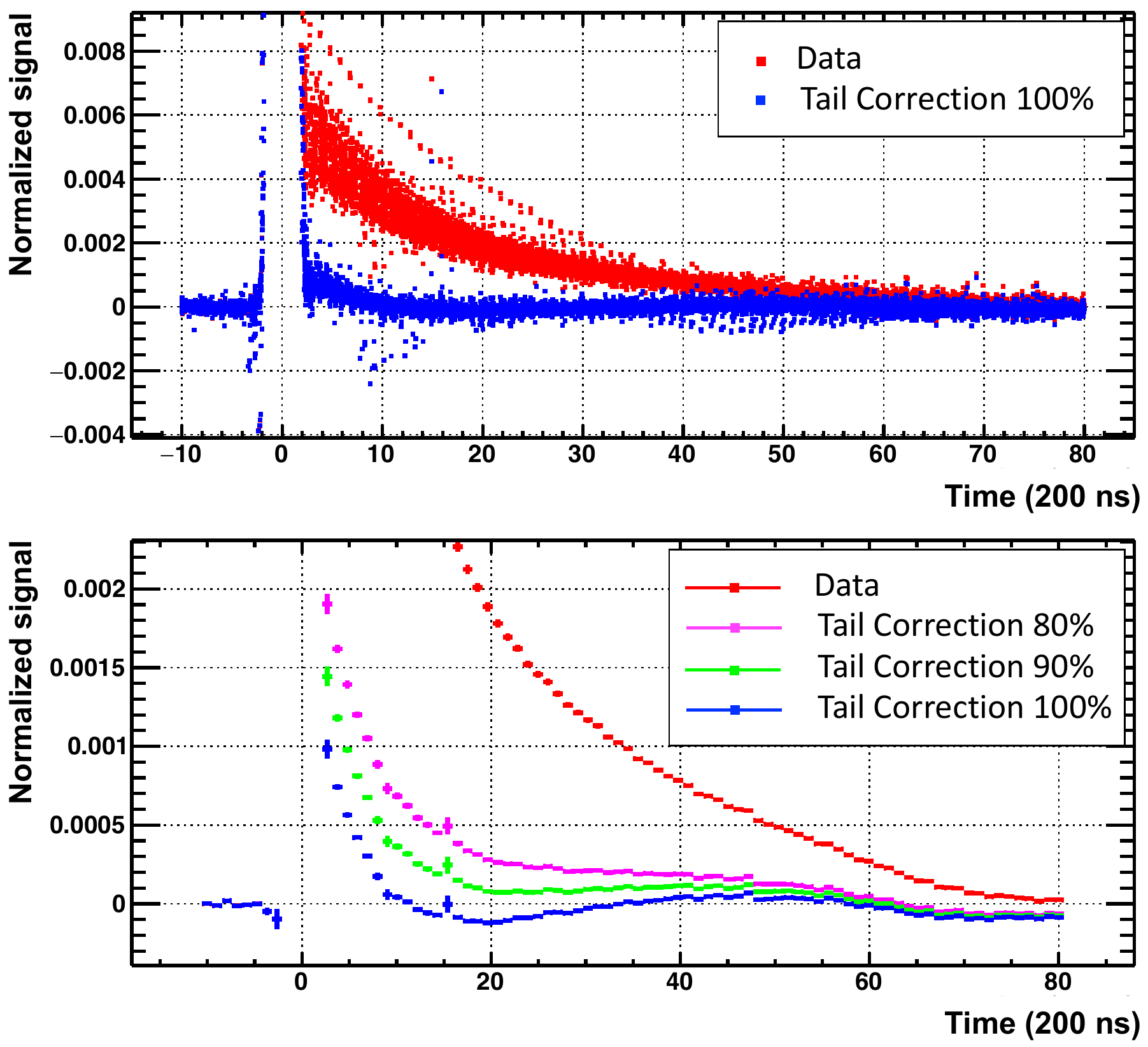}
	\caption{(Color online) Top: normalized ion tail before (red) and after (blue) correction for randomly selected pads. For each pad, the fitted tail parameters were used for the correction with \texttt{k0=100\%}. Bottom: averaged data with the corresponding corrections of 80\%, 90\% and 100\%. The error bars correspond to the RMS of entries for each time bin.}
	\label{fig:tailCorrAll}
\end{figure}

In \figref{fig:tailCorFract}, the exponential filter is applied on the averaged laser signal of a pad. The ion-tail fraction and ion-tail slope parameters, as obtained from the fit, were used for the correction. It can be seen that a 100\% correction (\texttt{k0=100\%}) slightly overcorrects the data, which is due to the bias of the sampled charge. \Figref{fig:tailCorrAll} (top) shows the normalized data and the 100\% correction for randomly selected pads of the nominal induction field setting, while the bottom panel shows the averaged data and the corresponding corrections. 

\begin{figure}[h!]
	\centering
	\includegraphics[width=0.7\linewidth]{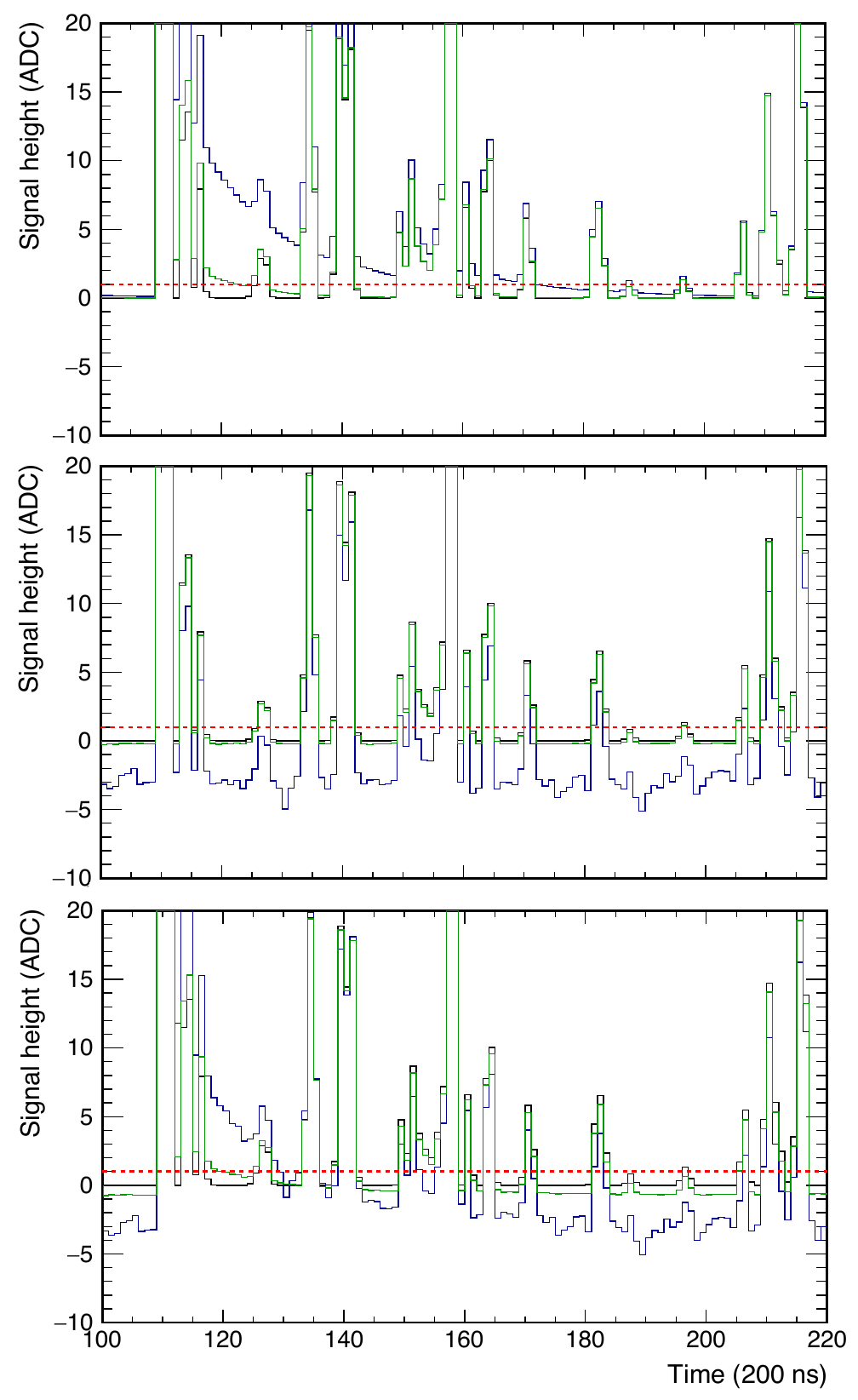}
	\caption{(Color online) Illustration of ion tail and common-mode correction of a simulated signal. The input signal is shown in black, the simulated effect (top: ion tail, middle: common-mode, bottom: ion tail and common-mode) in blue, and the corrected signal in green. Noise is not included in the pad signal for better visibility.}
	\label{fig:run3_baseline_corr}
\end{figure}

In \figref{fig:run3_baseline_corr} the simulation and correction of both effects are demonstrated. To show the effects more clearly, noise is not included. For the ion-tail correction, the ion-tail parameters of the respective pad were used, and a value of \texttt{k0=80\%} was chosen. For the common-mode correction, a two-iteration median correction was applied with \texttt{nPadsRandom=6}, \texttt{nPadsMin=4}, \texttt{Q\_thr1=2}, \texttt{Q\_thr2=2}.

\end{appendices}
\clearpage


\bibliographystyle{utphys}   
\bibliography{./biblio}

\end{document}